\documentclass[twocolumn,showpacs,amsmath,amssymb]{revtex4}

\usepackage{graphicx}
\usepackage{dcolumn}
\usepackage{bm}

\begin{document}
\title{An alternative to the plasma emission model: Particle-In-Cell, self-consistent 
electromagnetic wave emission simulations of solar type III radio bursts}
\author{David Tsiklauri}
\affiliation{Astronomy Unit, School of Mathematical Sciences, 
Queen Mary University of London, Mile End Road, London, E1 4NS, United Kingdom}
\date{\today}
\begin{abstract}
High-resolution (sub-Debye length grid size and 10000 particle species per cell), 
1.5D Particle-in-Cell, relativistic, fully electromagnetic simulations are used to model electromagnetic
wave emission generation in the context of solar type III radio bursts.
The model studies generation of electromagnetic waves by a super-thermal, hot beam of electrons injected into
a plasma thread that contains uniform longitudinal magnetic field and a parabolic density gradient.
In effect, a single magnetic line connecting Sun to earth is considered, for which five
cases are studied. (i) We find that the physical system without a beam is stable and only low amplitude level
electromagnetic drift waves (noise) are excited. (ii) The beam injection direction is controlled by
setting either longitudinal or oblique electron initial drift speed, i.e. by setting the beam pitch angle 
(the angle between the beam velocity vector and the direction of background magnetic field).
In the case of zero pitch angle i.e. when  $\vec v_b \cdot \vec E_\perp=0$,
the beam excites only electrostatic, standing waves, oscillating at local
plasma frequency, in the beam
injection spatial location, and only low level electromagnetic drift wave noise is also generated.
(iii) In the case of oblique beam pitch angles, i.e. when $\vec v_b \cdot \vec E_\perp \not =0$, 
again electrostatic waves with same properties are excited. However, now
the beam also generates the electromagnetic waves with the properties commensurate to type III radio bursts.
The latter is evidenced by the wavelet analysis of transverse electric field component, which
shows that as the beam moves to the regions of lower density and hence lower plasma frequency,
frequency of the electromagnetic waves drops accordingly. 
(iv) When the density gradient is removed,  an electron beam with an 
oblique pitch angle still generates the electromagnetic radiation. However, in the latter case no
frequency decrease is seen. 
(v) Since in most of the presented results the ratio of electron plasma
and cyclotron frequencies is close to unity near the beam injection location, 
in order to prove that the generated by the
non-zero pitch angle beam electromagnetic emission oscillates at the plasma frequency, we
also consider a case when the magnetic field (and the cyclotron frequency) is ten times
smaller. 
Within the limitations of the model, the study presents the first attempt to 
produce synthetic (simulated) dynamical spectrum of the type III radio bursts in the 
fully kinetic plasma model. The latter is based on 1.5D non-zero pitch angle (non-gyrotropic) electron beam,
that is an alternative to the plasma emission classical mechanism for which
two spatial dimensions are needed.
\end{abstract}	



\maketitle

\section{Introduction}
     
The type III solar radio bursts are known to be generated by the
super-thermal beams of electrons that travel away from the Sun
on open magnetic field lines \cite{1981ApJ...251..364L,1986ApJ...308..954L,1984A&A...141...30D}. The beams are likely to be
manifestations of magnetic reconnection which, in turn, is
driven by solar flares. However, flares can also drive 
dispersive Alfven waves which also can serve as a source of
super-thermal beams. In this work we do not focus on a question
what is an actual source of a beam. Instead, we consider a situation
when a hot $6\times10^6$K, super-thermal ($v_b=0.5c$) beam is injected into a 
cool $3\times10^5$K, Maxwellian plasma with 
parabolically decreasing density gradient,
along an open magnetic field line with $B=30$ gauss.
The latter mimics a magnetic field line that connects Sun to Earth.
There is large body of work done from the observational, modelling and
theoretical viewpoint. We refer the interested reader to 
appropriate reviews \cite{1987SoPh..111...89M,2002SSRv..101....1A,2008SoPh..253....3N,2008A&ARv..16....1P}
and to references in Ref. \cite{sp1}.
Also Introduction section of Ref. \cite{2010JGRA..11501101M} provides a good, critical overview of possible mechanisms which generate
the type III burst electromagnetic (EM) radiation. 
In brief, there are three categories of models of type III solar radio bursts:
(i) Quasilinear theory that uses kinetic Fokker-Planck type equation for describing
the dynamics of an electron beam, in conjunction with the spectral energy density evolutionary equations for  
Langmuir and ion-sound waves. In these models the spectral energy density of the Langmuir wave packets (that are excited by the
bump-on-tail unstable beam) travels along
the open magnetic field lines with a constant speed and this is despite the quasilinear relaxation 
(formation of a plateau in the electron distribution function). This implies some sort of beam marginal
stabilisation \cite{1968SvA....11..956K,1970AdA&A...7..147S,1972SoPh...24..444Z,1999SoPh..184..353M,2001A&A...375..629K,2002PhRvE..65f6408K}.
Some models also include EM emission into the quasilinear theory based on so-called 
drift approximation  
\cite{1988A&A...195..301H,1990A&A...229..216H,1999A&A...342..271H},
where nonlinear beam stabilisation during its propagation (so called free streaming) is based on Langmuir-ion acoustic wave coupling
via ponder-motive force and EM emission is prescribed by a power law  of the beam to ambient plasma number density ratio.
Such models can be used to construct and constrain the observed dynamical spectra physical parameters. 
(ii) Stochastic growth theory \cite{1992SoPh..139..147R,1992ApJ...387L.101R}, where density irregularities produce
a random growth, in such a way that Langmuir waves are generated stochastically
and quasilinear interactions within the Langmuir
clumps cause the beam to fluctuate about marginal stability. Such models can be also used for direct
comparison with the solar  type III bursts \cite{2008JGRA..11306104L}.
(iii) Direct, kinetic simulation 
approach of type III bursts \cite{2001JGR...10618693K,2005ApJ...622L.157S,2009ApJ...694..618R,2010JGRA..11501204U}  to this 
date used Particle-In-Cell (PIC) numerical method. These models mainly focus on the understanding of basic physics rather than
direct comparison with the observations. This is due to the size of simulation domain of the models being too small (only few 1000 Debye lengths which
is roughly $1 / 10^{10}$th of 1 AU).

In Ref.\cite{sp1} we have used
1.5D Vlasov-Maxwell simulations  to model EM emission generation in a fully 
self-consistent plasma kinetic model in the solar physics context. 
The simulations presented the generation of EM emission by 
the beam-generated Langmuir waves and Larmor drift instability
in a plasma thread that connects the Sun to Earth with 
the spatial scales compressed appropriately. We investigated 
the effects of spatial density gradients on the generation 
of EM radiation. In the case without an electron beam,
we found that the inhomogeneous plasma with a uniform background 
magnetic field directed transverse to the density
gradient is aperiodically unstable to the Larmor-drift instability. The latter
produced a novel effect of generation of EM emission at plasma frequency. 
The main results of Ref.\cite{sp1} can be summarised as following:
In the case without an electron beam, the induced perturbations
consist of two parts: (i) non-escaping Langmuir-type oscillations which are
localised in the regions of density inhomogeneity, and are highly filamentary, with the period of 
appearance of the filaments close to
electron plasma frequency in the dense regions; and (ii) escaping EM radiation with phase 
speeds close to the
speed of light. When we removed the density gradient (i.e. which then makes the plasma stable to Larmor-drift instability)
and a {\it low density}, super-thermal, hot beam is injected along the domain, 
in the direction perpendicular to the magnetic field (as in solar coronal magnetic traps which tend to
accelerate the particles in the direction perpendicular to the magnetic field \cite{2004A&A...419.1159K}),
the electron beam quasilinear relaxation generates non-escaping Langmuir type oscillations
which in turn generate escaping EM radiation.
We found that in the spatial location where the beam is injected, the standing waves, 
oscillating at the plasma frequency, are excited. It was suggested that these can be used to interpret
the horizontal strips (the narrow-band line emission) observed in some dynamical spectra \cite{2010A&A...515A...1A}.
We have also corroborated quasilinear theory predictions: (i) the electron free streaming and (ii) the beam long relaxation
time, in accord with the analytic expressions.
We also studied the interplay of Larmor-drift instability and the generation of EM emission by
the Langmuir waves
by considering {\it dense} electron beam in the Larmor-drift unstable (inhomogeneous) plasma. 
This enabled  us to study the deviations from the quasilinear theory.

In the present study we consider a situation that is more relevant to type III radio bursts.
The VALIS,  1.5D Vlasov-Maxwell code used in Ref.\cite{sp1} did not allow us to set the background magnetic
field along the physical domain (along $x$-axis) because it only solves for
$(E_x,E_y,0)$ and $(0,0,B_z)$ EM field components. For this reason the results of
Ref.\cite{sp1} were affected by the Larmor-drift instability. Thus, they were more applicable
to interpreting  the narrow-band line emission. Because, we considered spatially 1D situation in 
Ref.\cite{sp1} we had to set only one grid in the ignorable $y$-direction.
In the latter case, in the VALIS code a fluid equation is used to 
update the fluid velocity in the $y$- direction rather than 
Vlasov's equation. Since there is no pressure gradient in 
the $y$-direction, which is ignorable, the temperature plays no role.
Thus, it was not possible to set the electron beam velocity/momentum $y$-component.
In turn, the only option to set finite $k_{\perp}$ for the beam (which is a requirement
to excite EM wave, i.e. to couple the beam to EM emission) was to set small, 
finite perpendicular background magnetic field $B_{z0}$.
In effect, having finite $k_{\perp}$ for the beam means that
$\vec v_b \cdot \vec E_\perp \not =0$, i.e. the beam velocity/momentum vector has a projection on
the transverse EM component. Only in this case (in the case of non-zero pitch angle)
the beam can couple to EM wave \cite{abr88}.
Another way to look at the case considered in Ref.\cite{sp1} is to
realise that $\vec v_b \times \vec B_{z0} = \vec E$ (because the resistivity is zero and plasma beta is small,
electrons tend to be magnetised, 'frozen-into' plasma). In this case the latter vector product gives $ v_x B_{z0} = E_\perp=E_y$.
This is of course plausible in the solar corona, as the magnetic field has
all three components, but then the situation does not adequately describe type III bursts
in which the electron beams are believed to propagate along (not across!) the magnetic field lines.
For these reasons, now using EPOCH Particle-In-Cell (PIC) code, which can update all EM components,
and allows to set non-zero $B_{0x}$, we can (a) suppress the Larmor drift instability;
(b) set a finite electron beam velocity $y$-component at $t=0$ (hence to have $\vec v_b \cdot \vec E_\perp \not =0$)
which can readily excite EM waves; (c) hence consider the physical system that
adequately describes type III radio burst magnetic field geometry.

We would like to stress that the EM emission in our model is {\it different}
from the classical plasma emission mechanism. We elaborate on the difference
in the conclusions section.

The paper is organised as following: 
In sections III.a--III.c and III.e we consider the inhomogeneous density plasma, with
the density profile commensurate to type III radio bursts.
In Section III.a
we present the results of an equilibrium test run where the  initial conditions
described below are 
evolved   without imposing an electron beam;
In Section III.b we inject an electron beam strictly
along the background magnetic field, $B_{0x}$, with $p_{0x}=0.5\gamma m_e c$  ($\gamma=1/\sqrt{1-0.5^2}=1.155$ everywhere) and $p_{0y}=0$,
thus setting $\vec v_b \cdot \vec E_\perp=0$, in turn expecting that only electrostatic (ES) plasma
waves to be excited.
In Section III.c the electron beam is injected at an oblique angle
with $p_{0x}=0.5\gamma m_e c$ and $p_{0y}=0.5\gamma m_e c$,
thus setting $\vec v_b \cdot \vec E_\perp \not =0$, in turn expecting that EM waves to be excited.
For Section III.c we produce the synthetic (simulated) dynamical spectrum for the EM waves
by studying the behaviour of frequency of the EM emission generated by the beam 
as a function of time, which is expected to decrease, as the beam movies
into the regions of decreased density (hence decreased plasma frequency $\omega_{pe}$). 
Because as we will show below in our model the EM emission has 
frequency close to the plasma frequency, $\omega_{pe}$,
the fact that $\omega_{pe}\propto\sqrt{n_e}$ ensures that 
the  frequency of the EM emission decreases in time as the non-zero pitch angle 
electron beam moves towards the regions of progressively smaller background plasma
density.
In Section III.d we consider the situation identical to Section III.c, except
we set a uniform density to test the behaviour of frequency as function of time.  
In Section III.e we consider case similar to III.c but with 10 times weaker magnetic field
such that near the beam injection location $\omega_{ce}/\omega_{pe}(x=0)=0.094$, unlike in the rest of the paper where
$\omega_{ce}/\omega_{pe}(x=0)=0.94$ (which is commensurate to solar coronal conditions). 
This is to prove that the present study is indeed modelling 
the situation relevant for the
type III radio bursts, which emit near the plasma frequency, $\omega_{pe}$,
 rather than electron gyro-frequency, $\omega_{ce}$.

\section{The model}

We use EPOCH (Extendible Open PIC Collaboration) a multi-dimensional, fully electromagnetic, 
relativistic particle-in-cell code which was developed by 
EPSRC-funded 
Collaborative Computational Plasma Physics  (CCPP) consortium of 30 UK researchers. 
EPOCH uses second order accurate FDTD scheme to advance EM fields.
EPOCH's particle pusher is based on the Plasma-Simulation-Code (PSC) 
by Hartmut Ruhl, and is a Birdsall and
Landon type PIC scheme \cite{bl05} using Villasenor and Buneman current weighting. 
EPOCH uses a triangular shape function with the peak of the
triangle located at the position of the pseudo-particle and a width of twice the spatial grid length.
EPOCH utilises the Villasenor and Buneman \cite{1992CoPhC..69..306V} 
current calculating scheme which solves the additional equation
$\partial \rho / \partial t =\nabla \cdot \vec J$
to calculate the current at each time step. The main advantage of this scheme is that it conserves charge
on the grid rather than just globally conserving charge of the particles. This means that 
the solution of Poisson's equation is accurate to the machine precision, and when 
Poisson's equation is satisfied for $t = 0$ it remains
satisfied for all times. EPOCH has been thoroughly tested and benchmarked.

We use 1.5D version 
of the EPOCH code which means that we have one spatial component along $x$-axis 
and there are all three $V_x,V_y,V_z$ particle velocity components
present (for electrons, ions and electron beam).
Using these, the relativistic equations of motion are solved
for each individual plasma particle.
The code also solves Maxwell's equations, with self-consistent currents, using 
the full component set of EM fields
$E_x,E_y,E_z$  and $B_x,B_y,B_z$.
EPOCH uses un-normalised SI units, however, in order for our results
to be generic, we use the normalisation for the graphical
presentation of the results as follows.
 Distance and time are normalised to $c / \omega_{pe}$ and $\omega_{pe}^{-1}$, while
electric and magnetic fields to $\omega_{pe}c m_e /e$ and  $\omega_{pe} m_e /e$ respectively.
Note that when visualising the normalised results we use   $n_0 =10^{14}$ m$^{-3}$ in the
densest parts of the domain, which are located at the leftmost and rightmost edges of the
simulation domain
(i.e. fix $\omega_{pe} = 5.64 \times 10^8$ Hz radian in the densest regions).
Here $\omega_{pe} = \sqrt{n_e e^2/(\varepsilon_0 m_e)}$ is the electron plasma frequency,
$n_\alpha$ is the number density of species $\alpha$ and all other symbols have their usual meaning.
We intend to consider a single plasma thread (i.e. to use 1.5D geometry), therefore
space component considered here has $x=65000$ grid points, with the grid 
size is  $\lambda_D/2$  for the two long runs (Sections III.c and III.e) making maximal value for
$x$, $x_{max}=231.170 c / \omega_{pe}$, while in the short runs 
(Sections III.a-b and III.d) the grid 
size is  $\lambda_D/4$  yielding the maximal value for
$x$, $x_{max}=115.585 c / \omega_{pe}$. 
Here $\lambda_D = v_{th,e}/ \omega_{pe}$
is the Debye length ($v_{th,e}=\sqrt{k T/m_e}$ is electron thermal speed).
Since we would like to resolve full plasma kinetics,
our choice of the grid size is 2 to 4 times better than in  Ref.\cite{sp1}
where only
spatial grid size of $1 \lambda_D$ was used.
Thus, the presented results can be regarded as
high resolution (sub-Debye length scale), guaranteeing a superior capture of kinetic
physics.     

We do not fix  plasma
number density and hence $\omega_{pe}$ deliberately, because we wish our results to stay general.
We demonstrate this  on the 
following example, if we set plasma number density to
 $n_0 =10^{14}$ m$^{-3}$ (i.e. fix $\omega_{pe} = 5.64 \times 10^8$ Hz radian),
this sets Debye length at $\lambda_D = 3.78\times10^{-3}$m=$7.11\times10^{-3} c / \omega_{pe}$ (using $T=3\times10^5$K).
If we set plasma number density to
 $n_0 =10^{-5}$ m$^{-3}$,
this sets Debye length at $\lambda_D = 1.20\times10^{7}$m=$7.11\times10^{-3} c / \omega_{pe}$.
Thus, appropriately adjusting plasma number density
$n_0$, physical domain can have arbitrary size e.g. Sun-earth distance (but then
unrealistically low density has to be assumed). 
Background plasma in our numerical simulation is assumed to be 
Maxwellian, cool $T=3\times10^5=const$  with 
parabolically decreasing density gradient,
along the uniform magnetic field line.
The latter mimics a field line that connects Sun to Earth.
The only physical parameters that should be regarded as fixed are the temperatures of the
background plasma, that of the electron beam and magnetic field ($B_{0x}=30$ gauss). 
These are set to, plausible for the
type III bursts values, $T=3\times10^5$K for the background plasma and $T_b=6\times10^6$K
for the beam. This fixes respective electron thermal speeds to 
$v_{th,e}=7.11\times 10^{-3}c$ and $v_{th,b}=3.18\times 10^{-2}c$.
If an attempt is made to interpret some type III burst
observations, one should keep in mind whilst density and hence
$\omega_{pe}$ can be regarded as variable (arbitrary), $T$, $T_b$ and $B_{0x}$ are fixed (model specific).
Ref.\cite{2009ApJ...706L.265C} have shown that
parabolic density profile $n_e(r) $
describes the electron number density to a good approximation
within few solar radii, $ R_{\odot}$.
Generally, $n_e(r)\propto r^{-2}$
plasma number density profile can be well understood based on
conservation of mass for a spherically symmetric constant speed outflow
such as Parker's solar wind solution. 
For large radii, $r \gg R_{\odot}$, practically all models predict
$n_e(r)\propto r^{-\delta}$ scaling with $\delta$ being close to two,
e.g. 2.16 in Ref.\cite{1999A&A...348..614M} or 2.19 in Ref.\cite{1998SoPh..181..429R}.
Therefore, to a good approximation, we use following density profile
for the background electrons (and ions):
\begin{equation}
n_0(x)=\left((x-x_{max}/2)/(x_{max}/2 + n_{+} )\right)^2+n_{-},
\end{equation}
where $n_0(x)$ is the normalised plasma number density, such that
for the left and right edges of the simulation domain, $x=0$ and $x=x_{max}$,
$n_0(0)=n_0(x_{max})=1$, while in the middle, $x=x_{max}/2$,  the parameters 
$n_{+}=(x_{max}/2) \times (1-\sqrt{1-n_{-}})/\sqrt{1-n_{-}}$ and
$n_{-}=10^{-8}$ where chosen such that $n_0(x)$ drops $10^{-8}$ times compared to the
edges. 
This density profile effectively 
mimics a factor of $10^8$ density drop from the corona  $n_0 =10^{14}$ m$^{-3}$
to $n_{AU} =10^{6}$ m$^{-3}$ at 1 AU. Because numerically 
most precisely implementable boundary conditions are the periodic ones, 
this density profile
represents mirror-periodic situation when the domain size is effectively doubled, i.e.
at $n_0(x=0)=n_0(x=x_{max})=1$ while  $n_0(x=x_{max}/2)=10^{-8}$.
This way "useful" or "working" part of the simulation domain is 
$0 \leq x \leq x_{max}/2$.
When cases with the beam are considered we set its following 
density profile:
\begin{equation}
n_b(x)= {\tilde n_b}e^{-[ ({x-x_{max}/25})/({x_{max}/40}) ]^8 }
\end{equation}
which means that the beam is injected at $x=x_{max}/25$ and its 
full width at half maximum (FWHM) is  $\approx x_{max}/20$ (see Figure 4(c) dashed curve).

As, we impose background magnetic field $B_{0x} = 30=const$ gauss along
$x$-axis, plasma beta in this study, based on the above parameters, is set 
to $\beta= 2 (v_{th,i}/c)^2(\omega_{pi}/\omega_{ci})^2 = 
n_0(0)k T /(B_0^2/(2\mu_0))=1.16 \times10^{-2}$ at $x=0$. 
It should be noted that the pressure balance in the initial conditions
is not kept. There are two reasons for this: (i) solar wind is not in "pressure balance" 
and it is a continually expanding solar atmosphere solution; (ii) plasma beta is small
therefore it is not crucial to keep {\it thermodynamic} pressure in balance
(because its effect on total balance is negligible) and
the initial background density stays intact throughout the simulation time (see e.g. Figure 4(c), 
thick solid curve).

EPOCH code allows to set an arbitrary number of plasma particle species.
Thus, since we intend to study spatially localised electron beam 
injected into the inhomogeneous or homogeneous Maxwellian electron-ion plasma,
we solve for three plasma species electrons, ions and the electron beam.
The dynamics of the three species, which mutually interact via EM interaction,
can be studied independently in the numerical code.
Velocity distribution function for electrons and ions is always set to 
\begin{equation} 
f_{e,i}=e^{-(p_x^2+p_y^2+p_z^2)/(2 m_{e,i} k T) },
\end{equation}
where the momenta components, $p_{x},p_{y},p_{z}$, include the correct electron and ion masses which are different by the 
usual factor of $m_i/m_e=1836$.
When cases with the beam are considered we set its following distribution
\begin{equation}
f_{b}= {\tilde n_b} e^{-((p_x-p_{x0})^2+(p_y-p_{y0})^2+p_z^2)/(2 m_{e,i} k T) }.
\end{equation}
where ${\tilde n_b}$ is normalised beam number density (${\tilde n_b}=n_b/n_{e0}$ ) and it is 
${\tilde n_b}= 10^{-3}$ throughout this study.
This choice is on the limit of available current computational facilities used --
64  Dual Quad-core Xeon $= 64 \times 8 = 512$ processor cores with 4 Tb of RAM. 
Typical run takes 28 hours on 512 processors.
We have used  $6.5 \times 10^8$ electrons, $6.5 \times 10^8$ ions and $6.5 \times 10^5$
 beam electrons giving a total of $1.30065 \times 10^9$
particles in the simulation. with 65000 spatial grid points this means that we sample
electron and ion phase space very well with 10000 per simulation cell.
With ${\tilde n_b}= 10^{-3}$ this means that globally (on average) we only have $10000\times 10^{-3}=10$
electrons to represent the beam. Thus, we cannot consider a realistic ${\tilde n_b}= 10^{-5}-10^{-7}$
commensurate to type III bursts. However as can be seen from Figure 4(c), dashed curve, the beam is quite
localised, about 1/10th  of the domain length (i.e. twice the FWHM $\approx 2 x_{max}/20 =x_{max}/10$). Thus, in reality $6.5 \times 10^5$
beam electrons are loaded into $65000/10=6500$ cells providing reasonably good $6.5 \times 10^5/6500=100$ beam particles per cell.

\section{Results}

Below we present numerical simulation results for the four runs.
We use the beam injection initial momentum components $p_{x0}$ and  $p_{y0}$
to control what type of waves can be excited by the beam, as well as study the effect of the background
plasma density gradient on the generation and properties of EM waves.

\subsection{Inhomogeneous plasma without an electron beam}

\begin{figure*}    
   \centerline{\hspace*{0.015\textwidth}
               \includegraphics[width=0.515\textwidth,clip=]{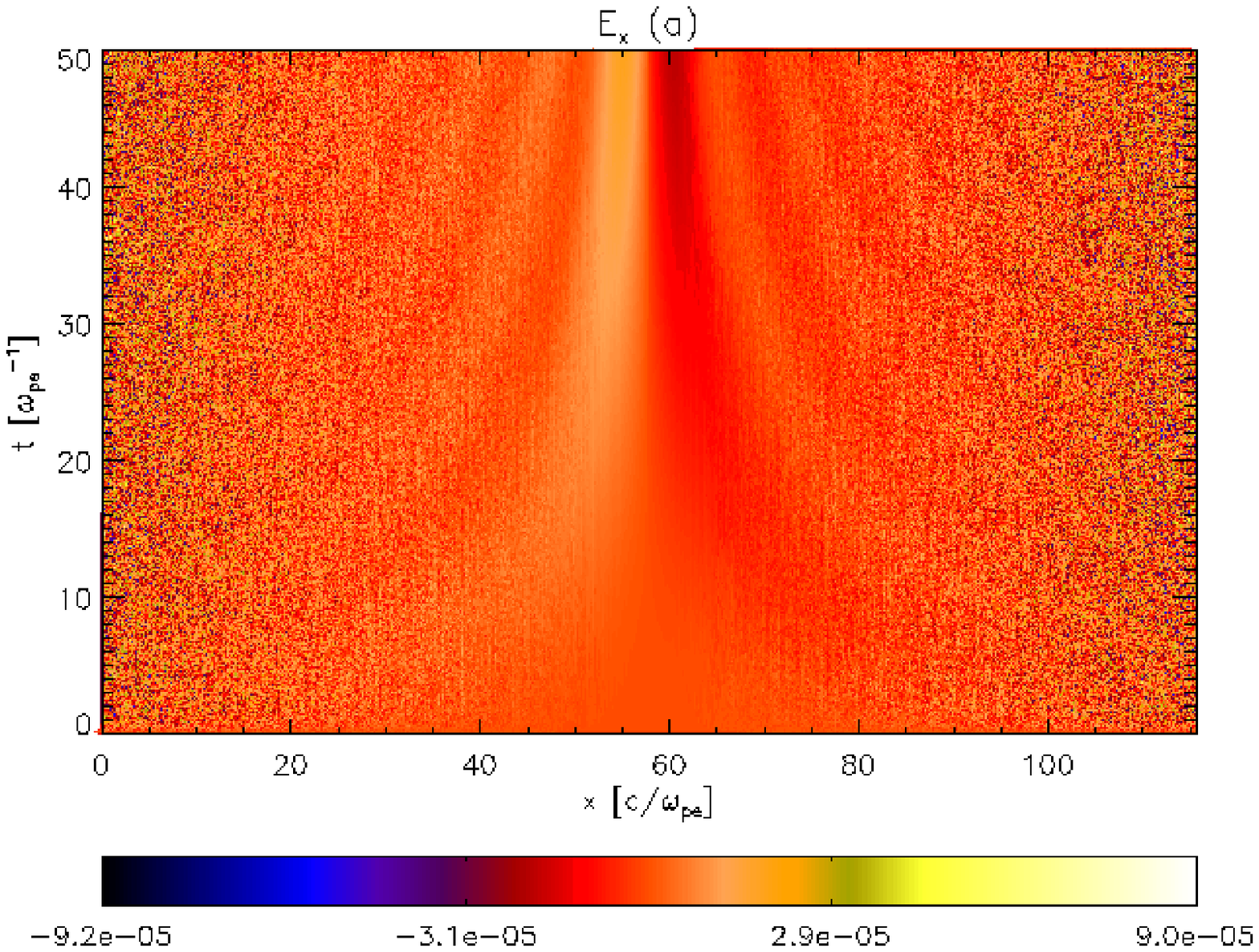}
               \hspace*{-0.03\textwidth}
               \includegraphics[width=0.515\textwidth,clip=]{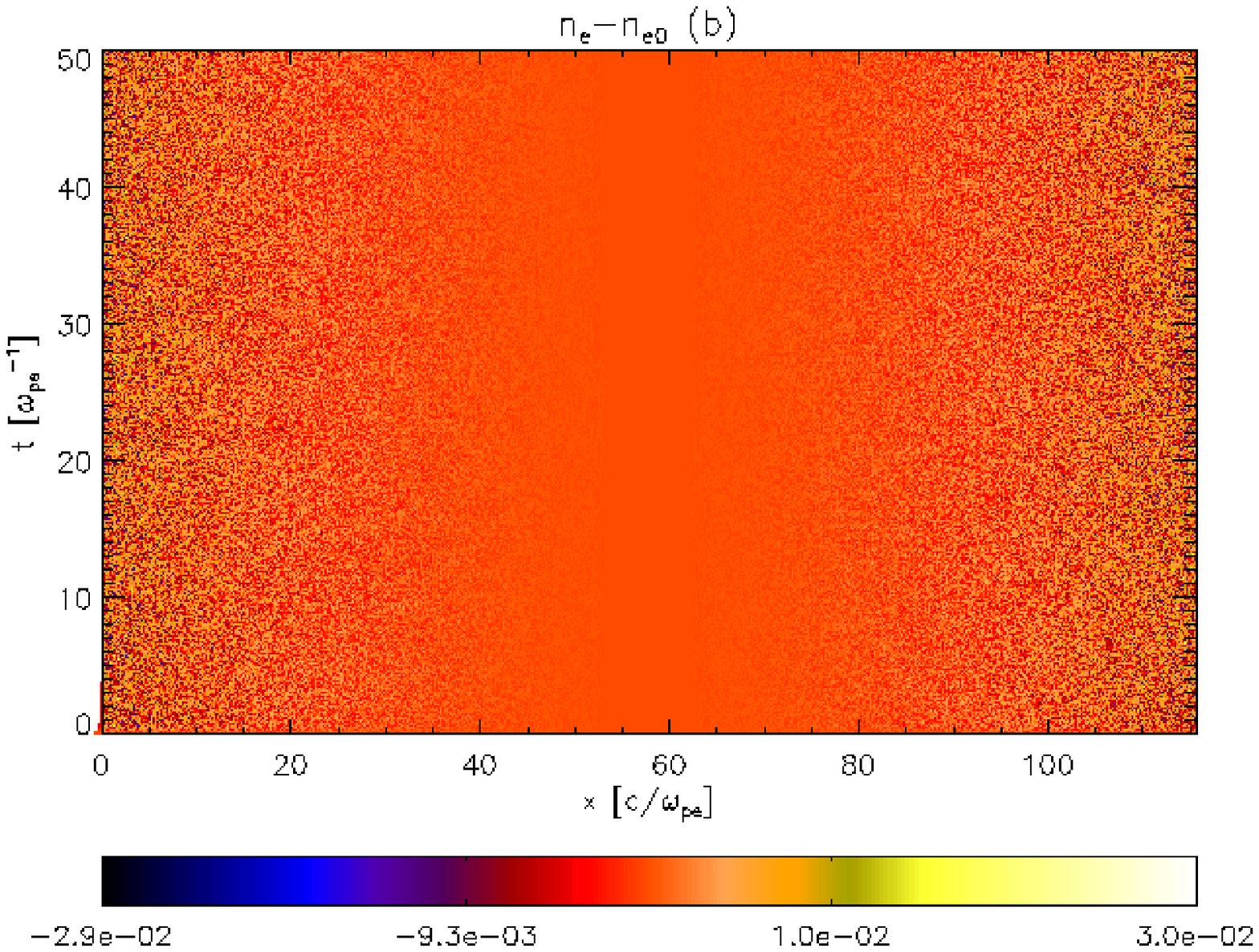}
              }
   \centerline{\hspace*{0.015\textwidth}
               \includegraphics[width=0.515\textwidth,clip=]{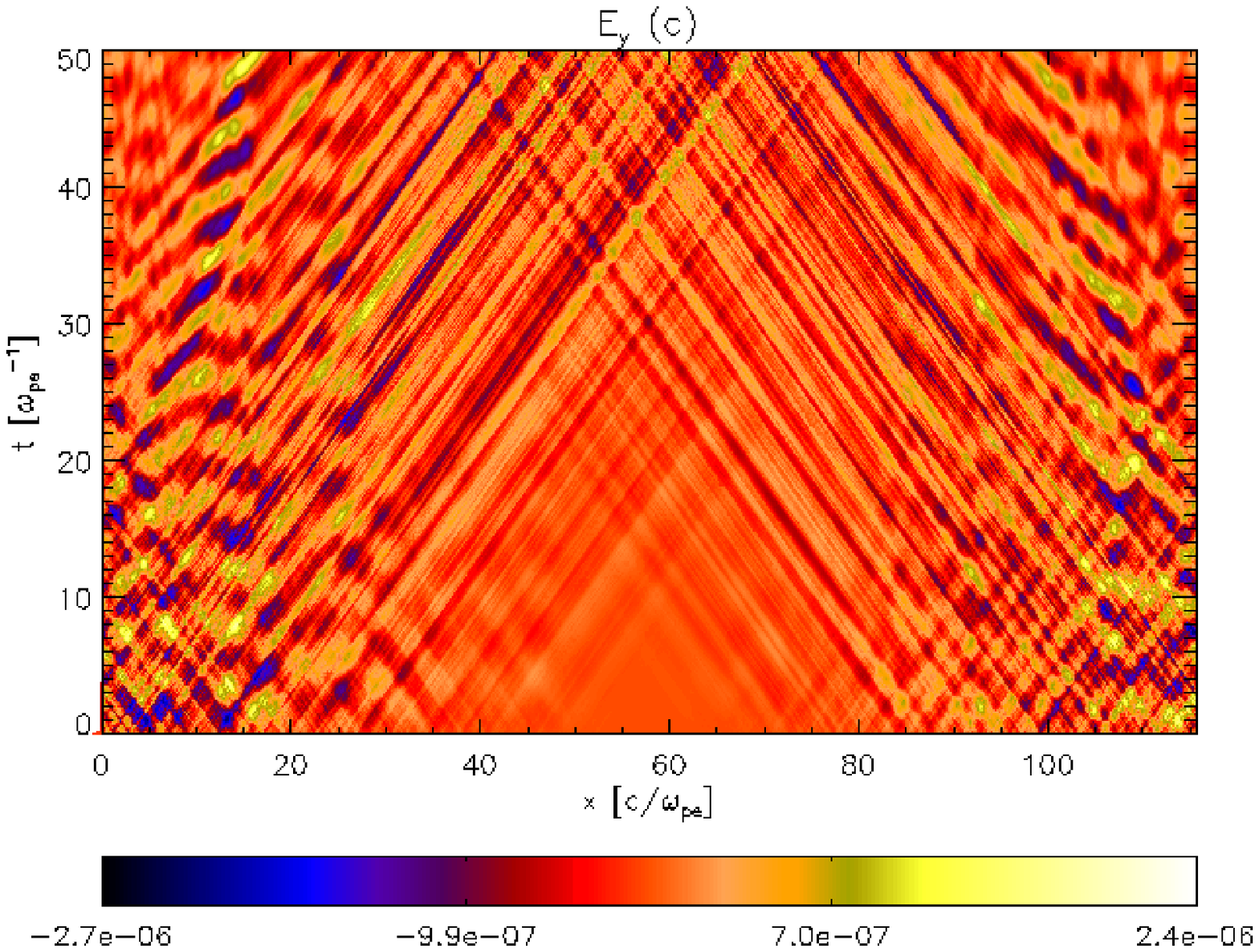}
               \hspace*{-0.03\textwidth}
               \includegraphics[width=0.515\textwidth,clip=]{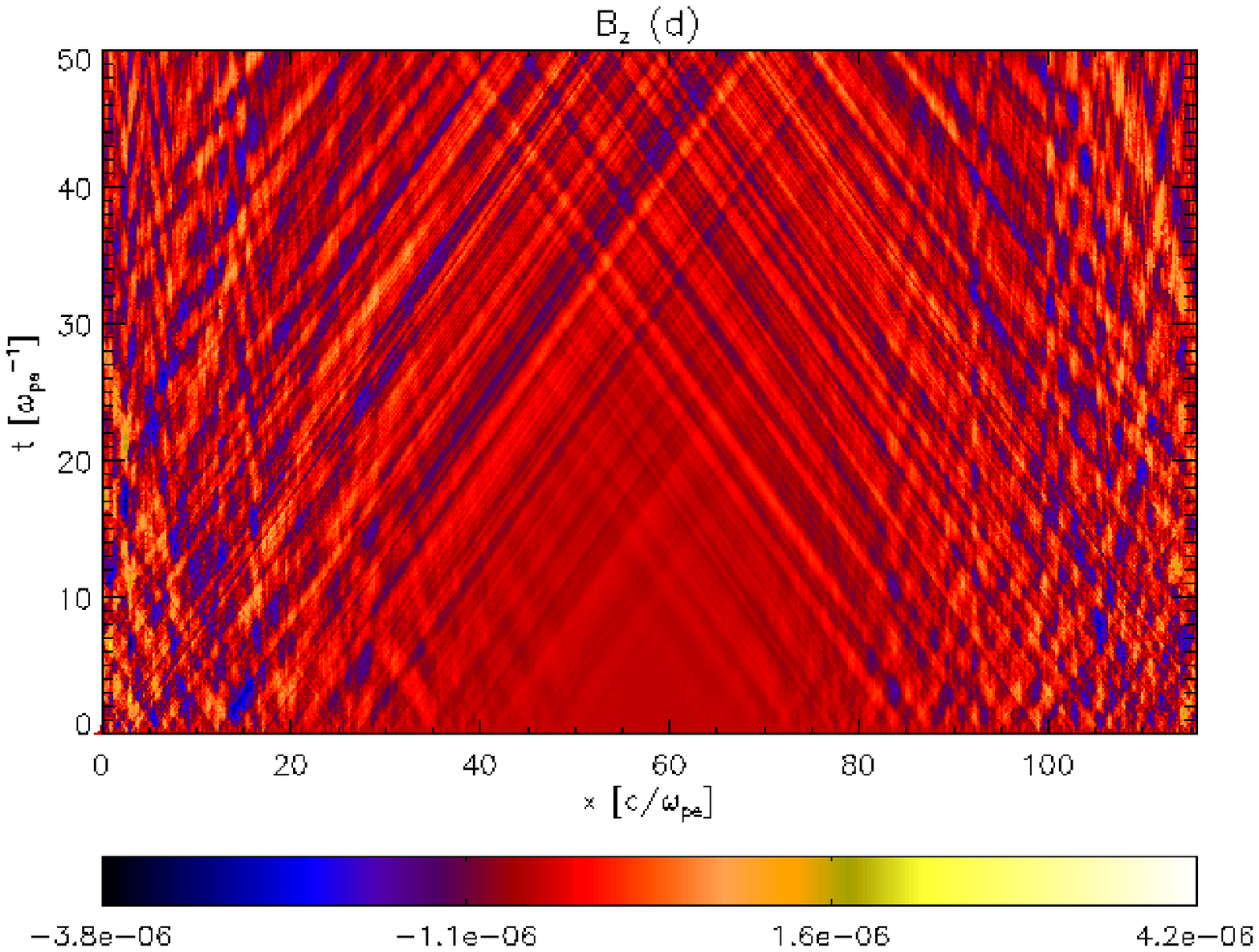}
              }
	      
	      \centerline{\hspace*{0.015\textwidth}
               \includegraphics[width=0.515\textwidth,clip=]{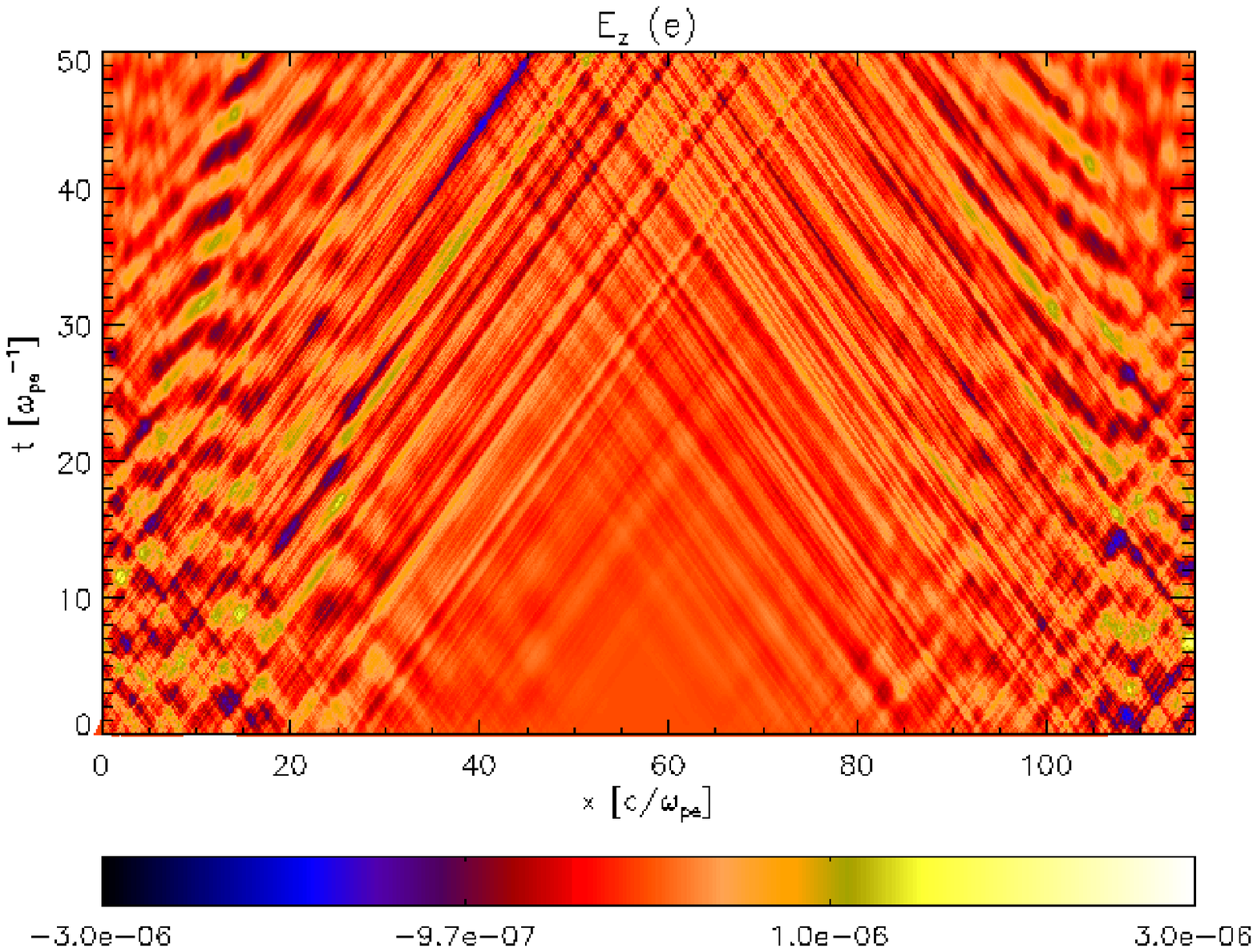}
               \hspace*{-0.03\textwidth}
               \includegraphics[width=0.515\textwidth,clip=]{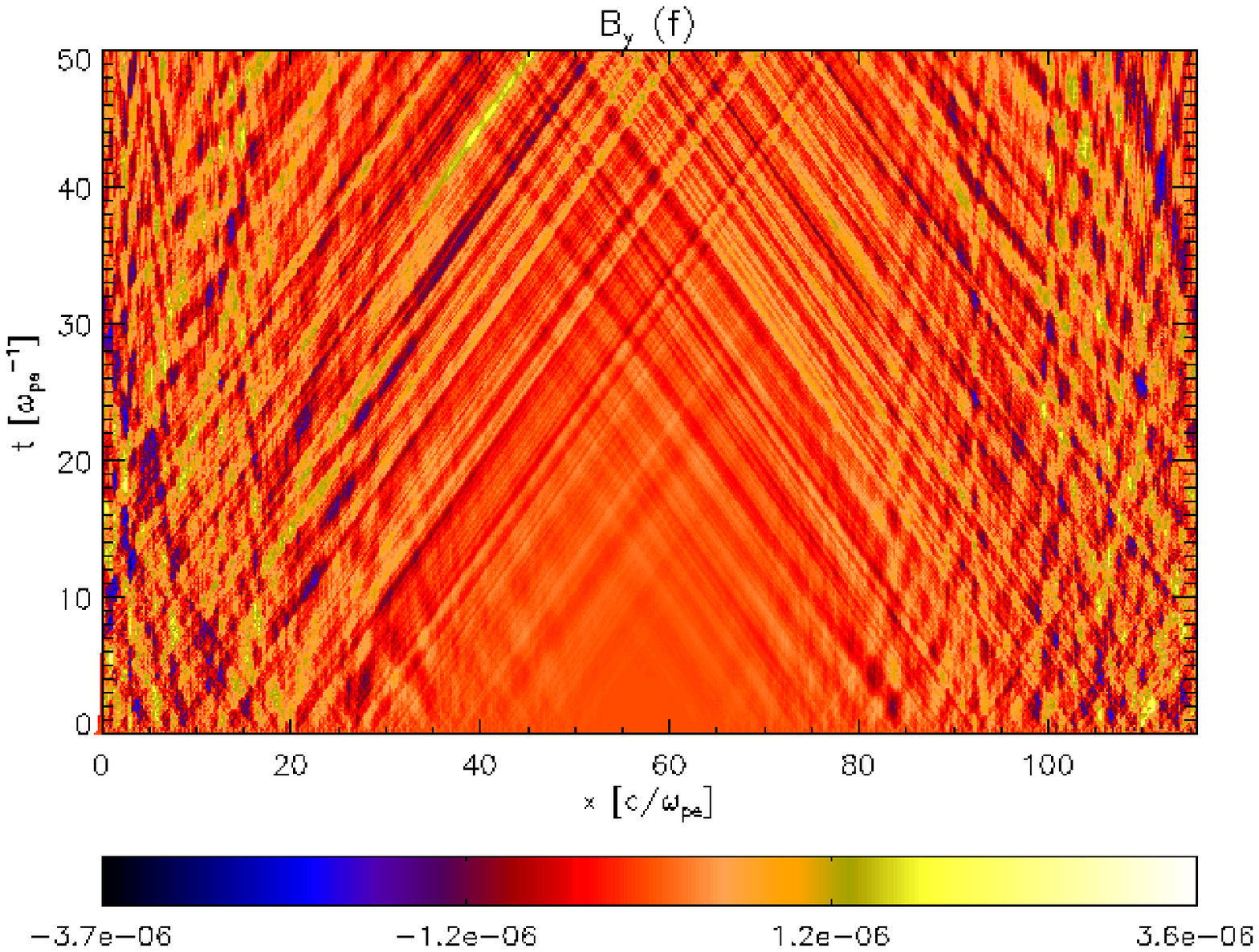}
              }
\caption{Time-distance plots for: (a) $E_x$,  (b) $n_e -n_{e0}$, (c) $E_y$, (d) $B_z$, (e) $E_z$ and (f) $B_y$.
This figure pertains to Section III.a.}
   \end{figure*}

In this section we present an equilibrium test run where the above described initial conditions are 
evolved for $50\omega_{pe}^{-1}$  without imposing an electron beam.
The results are presented in Figure 1 where we show
time-distance plots for the electrostatic (longitudinal to both background magnetic field and
density gradient) electric field $E_x$ 1(a) and associated density perturbation,
$n_e-n_{e0}$ 1(b); and two components of the transverse electromagnetic fields:
($E_y$ 1(c), $B_z$ 1(d)) and ($E_z$ 1(e), $B_y$ 1(f)).
We gather from Figures 1(a) and 1(b) that only low level noise (with amplitudes $\sim 9\times 10^{-5}$ for $E_x$ and  
$\sim 3\times 10^{-2}$ for $n_e-n_{e0}$)
is generated. This can be attributed to so called "shot noise" that is normally present in PIC simulations.
Figures 1(c)--1(f) demonstrate that in the electromagnetic emission component also low level (with amplitudes few $10^{-6}$)
drift EM wave noise is  generated. This can be evidenced by the fact that
the slope of bright and dark strips is roughly the speed of light. Hence the perturbations are travelling with the speed of light.
Note that the perturbations are generated in all parts of the density gradient but they are more prominent in the
densest parts of the simulation domain, because their amplitudes are also expected to be largest there.
No regular perturbations of longitudinal magnetic field ($B_x-B_{0x}$) are found (not shown here).
Thus, we conclude that the equilibrium without the electron beam is fairly stable (apart from the low level EM drift wave noise).

\subsection{Inhomogeneous plasma with electron beam injected  along the magnetic field ($\theta =0^{\circ}$)}

\begin{figure*}    
   \centerline{\hspace*{0.015\textwidth}
               \includegraphics[width=0.515\textwidth,clip=]{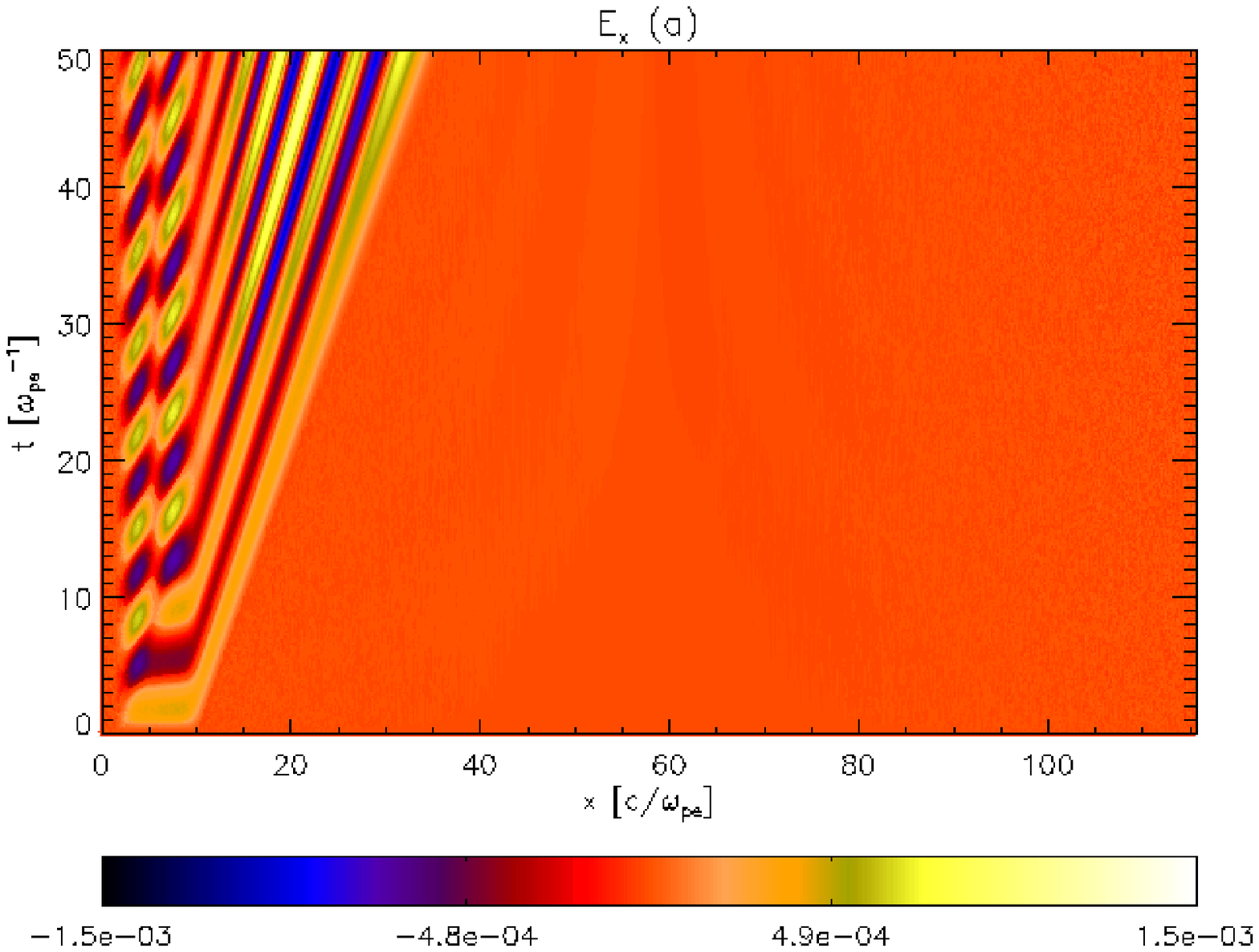}
               \hspace*{-0.03\textwidth}
               \includegraphics[width=0.515\textwidth,clip=]{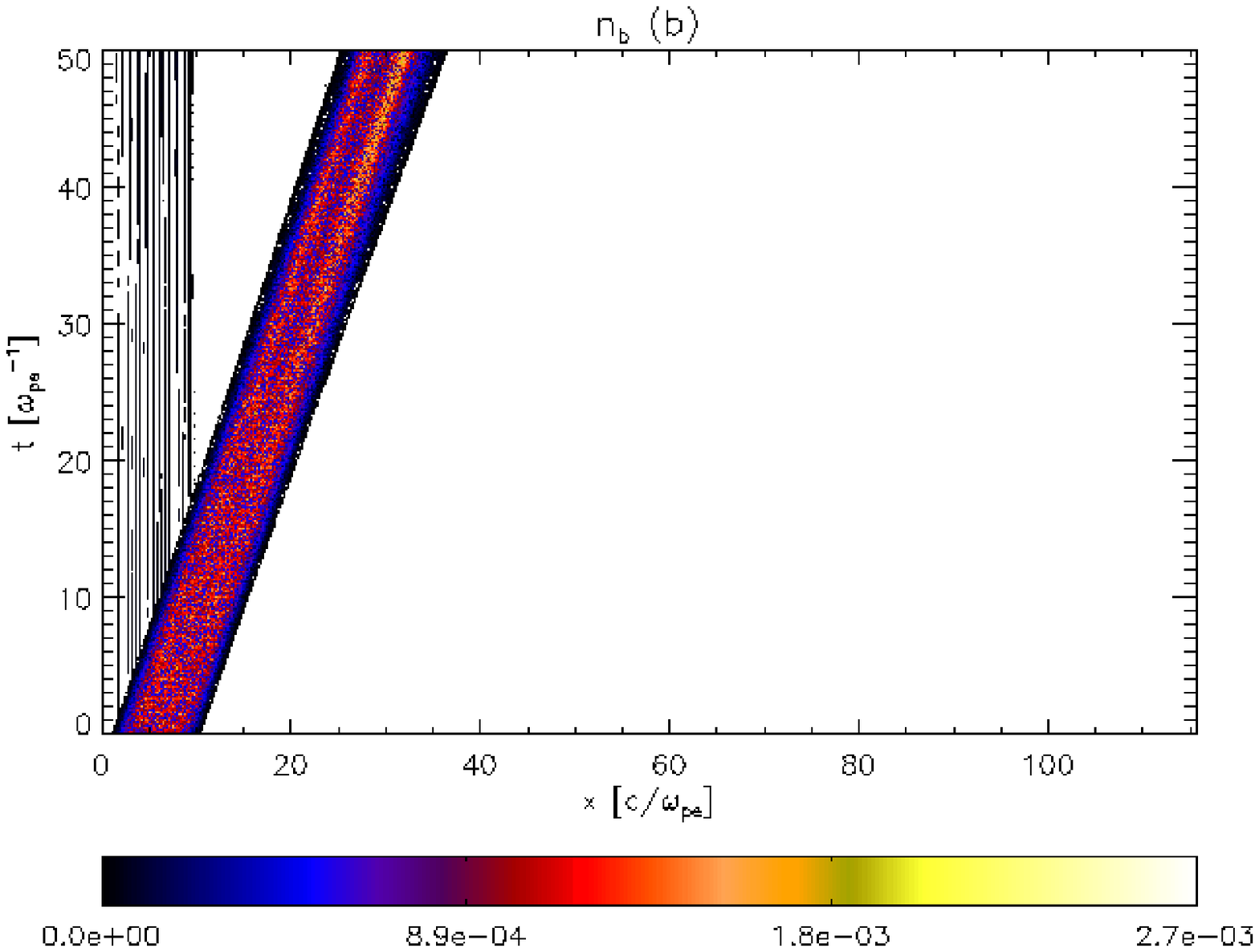}
              }
   \centerline{\hspace*{0.015\textwidth}
               \includegraphics[width=0.515\textwidth,clip=]{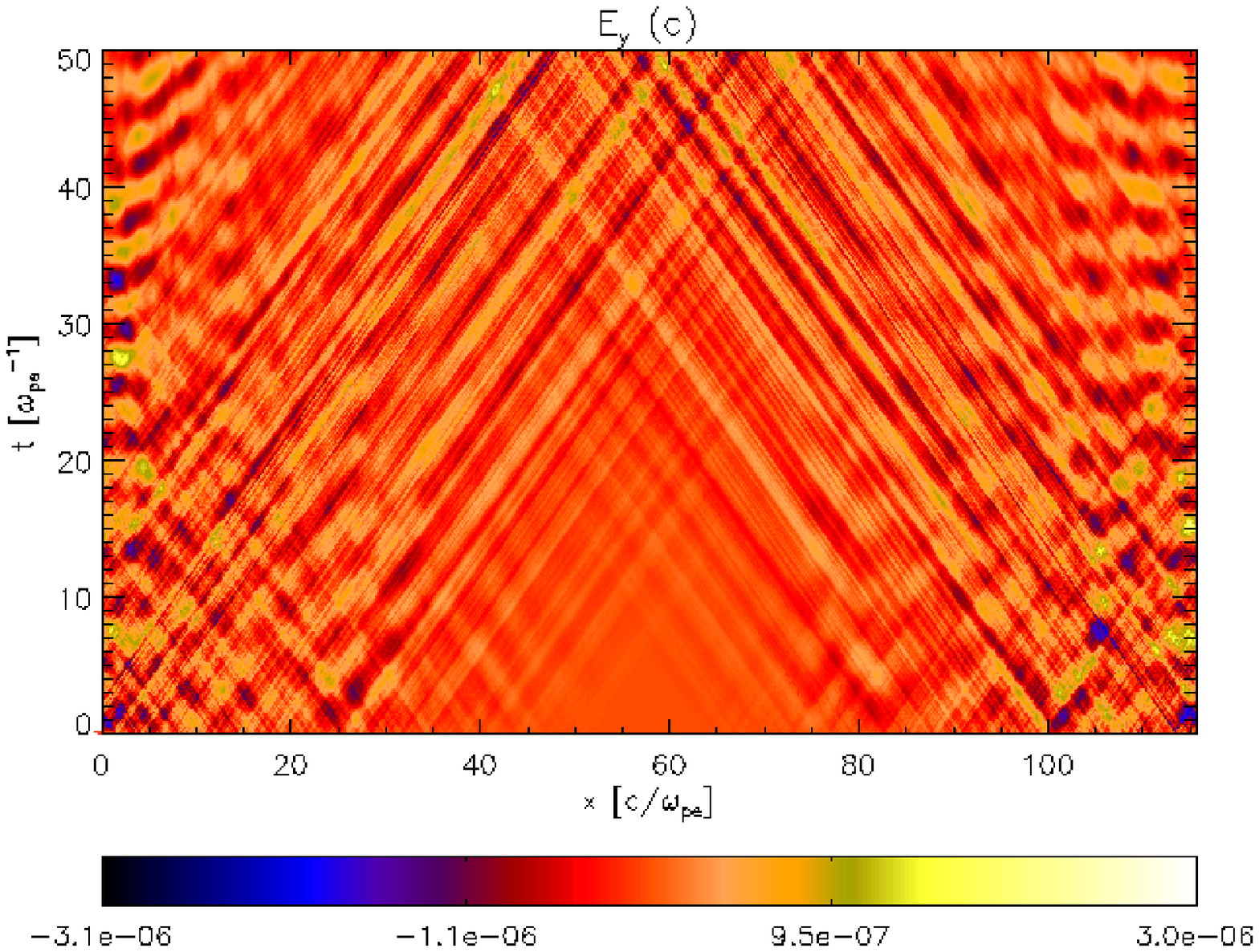}
               \hspace*{-0.03\textwidth}
               \includegraphics[width=0.515\textwidth,clip=]{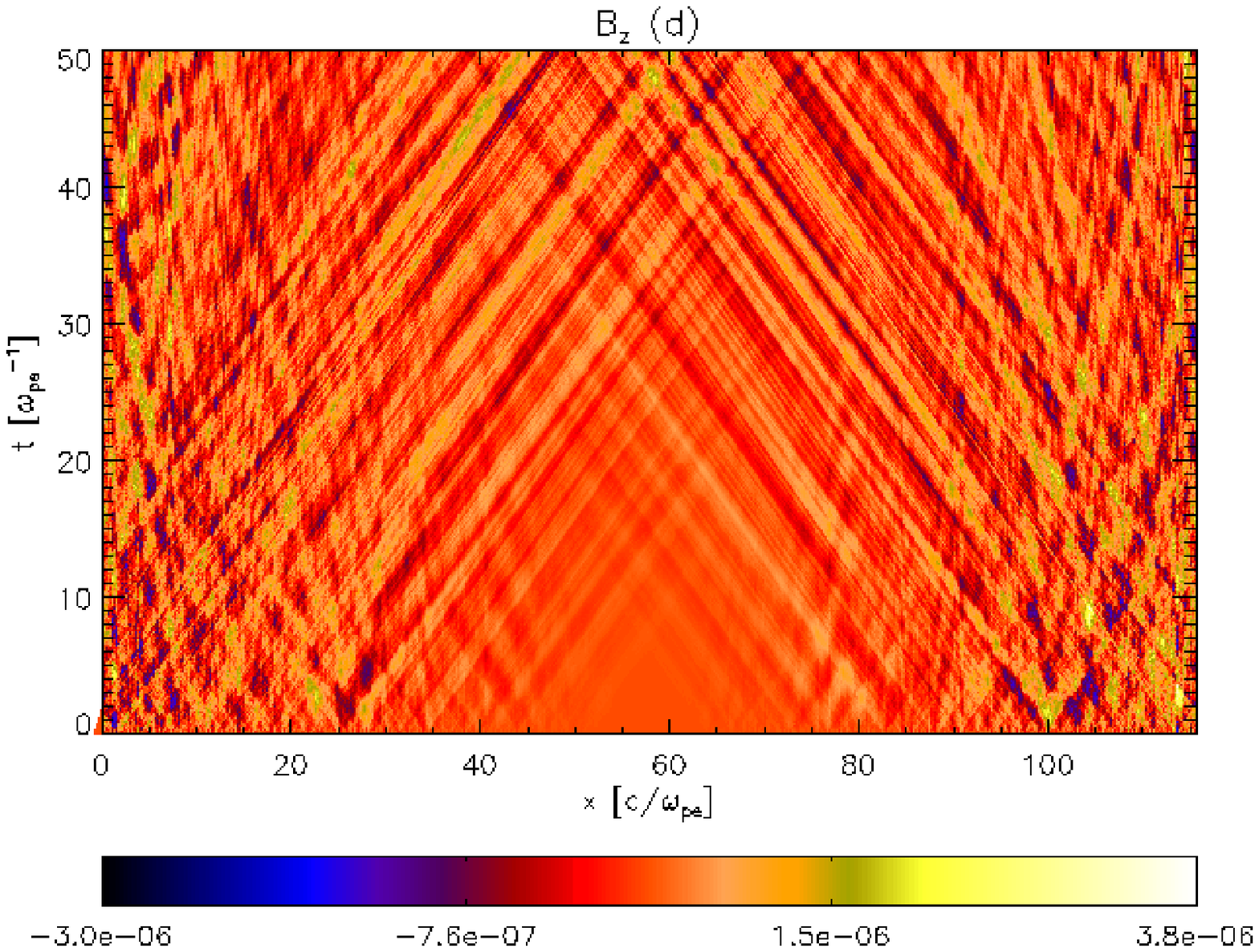}
              }
	      
	      \centerline{\hspace*{0.015\textwidth}
               \includegraphics[width=0.515\textwidth,clip=]{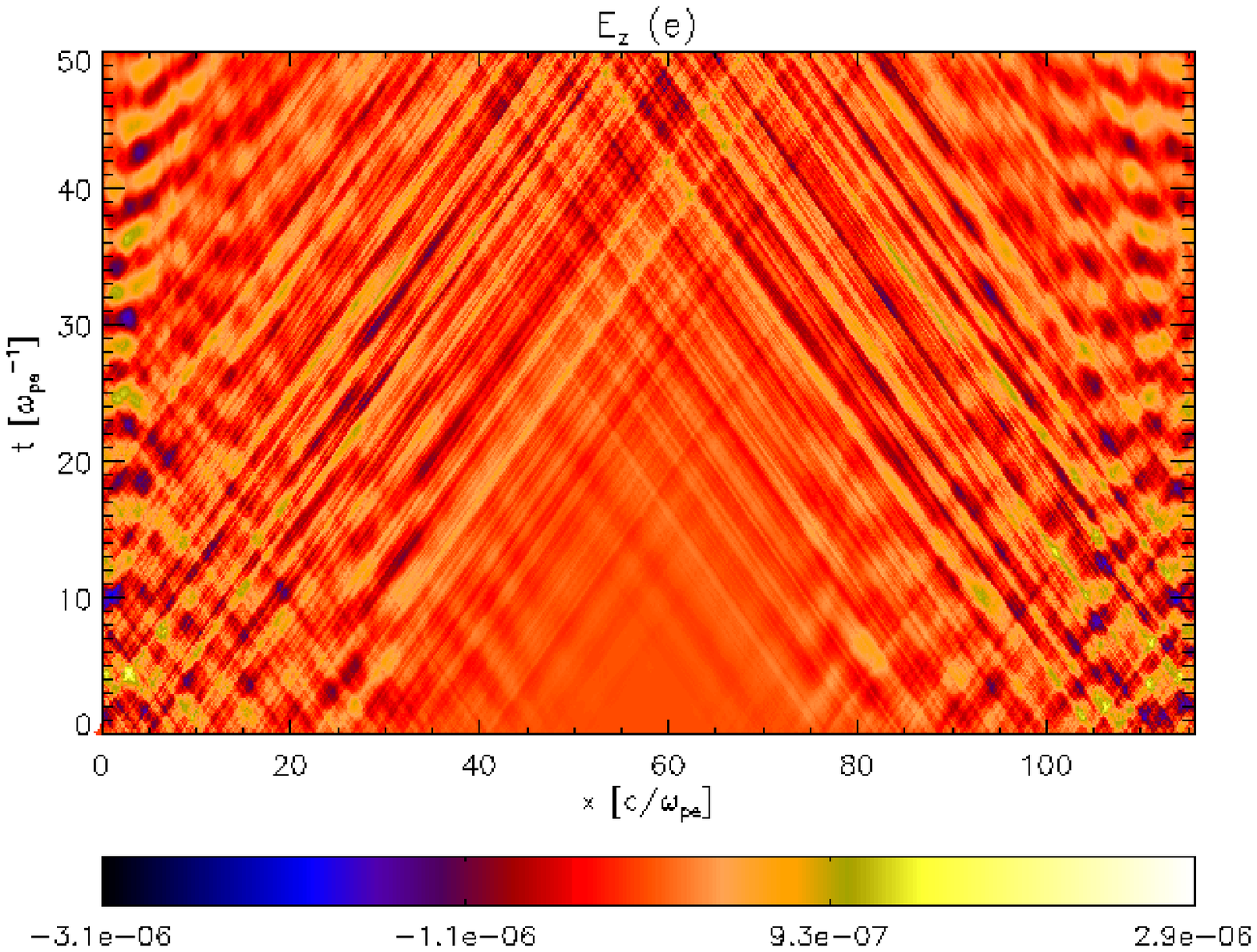}
               \hspace*{-0.03\textwidth}
               \includegraphics[width=0.515\textwidth,clip=]{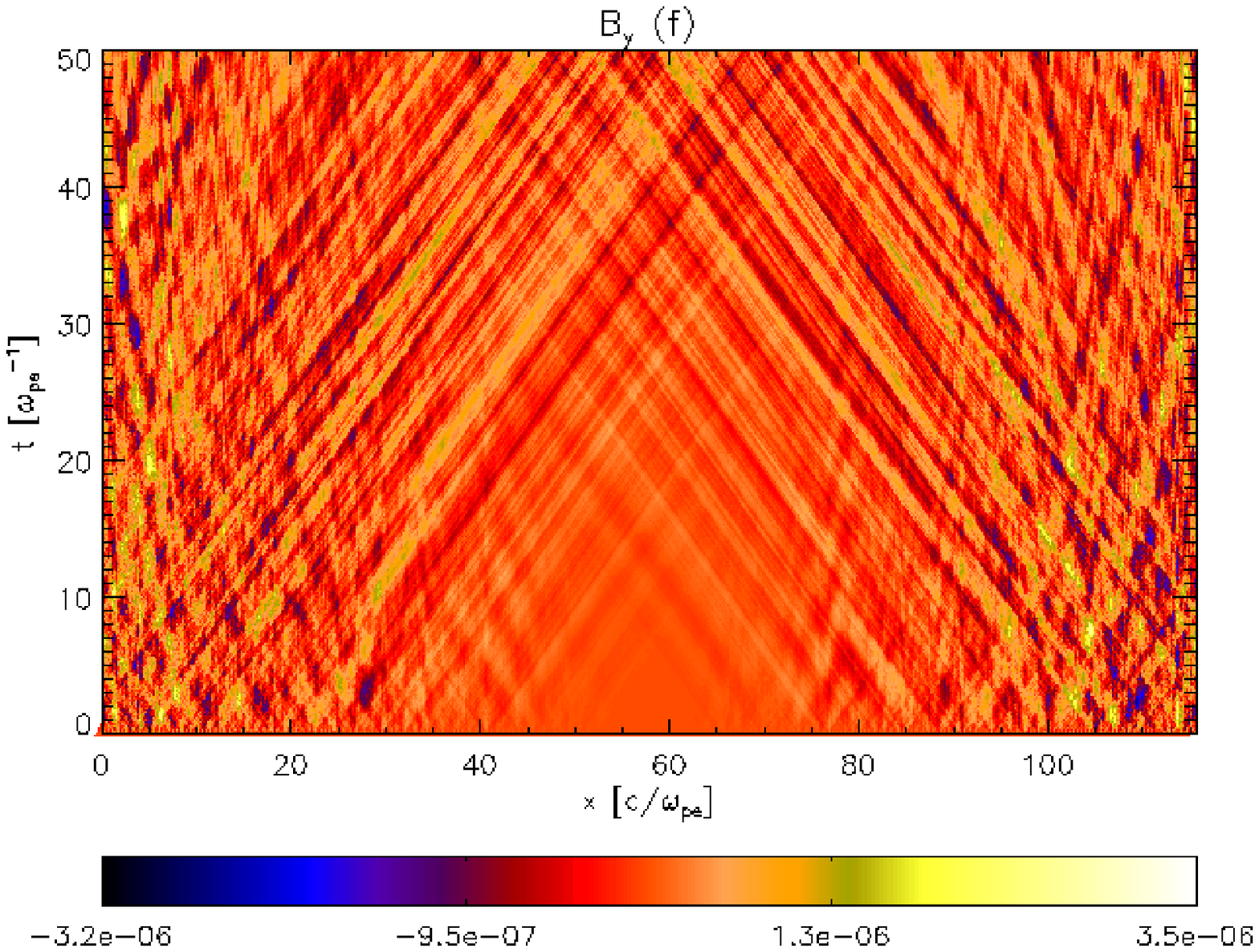}
              }
\caption{Time-distance plots for: (a) $E_x$,  (b) $n_b$, the electron beam number density, (c) $E_y$, (d) $B_z$, (e) $E_z$ and (f) $B_y$.
Here the beam pitch angle $\theta=0$. This figure pertains to Section III.b.}
   \end{figure*}
   
In this section we present the results when we inject the electron beam  along the background magnetic field ($\theta =0^{\circ}$), where $\theta$ is
the beam pitch angle (the angle between the initial beam velocity/momentum vector and the direction of background magnetic field).
Here in Eq.(4) at $t=0$ we set  $p_{0x}=0.5\gamma m_e c$ and $p_{0y}=0$.
Note that in all cases with the injected beam we solve the initial value problem, i.e.
electron beam {\it initial drift momentum is applied only at} $t=0$ -- we do not re-inject the beam at every time step.
The results are shown in Figure 2 where physical quantities shown
are similar to that of Figure 1 except for 1(b) where instead we now present the time-distance plot for the electron beam, $n_b$.
The reason why we have chosen to trace the dynamics of $n_b$ rather than full electron number density perturbation (background
 electron population plus the
beam) is because as it was show in Section 2.3 of Ref.\cite{sp1}, when the beam is relatively dense ($n_b/n_e \approx$ few $10^{-2}$),
the electron number density perturbation is dominated by a wake created by the beam. 
In Ref.\cite{sp1} it was shown that when an electron beam, with the properties similar to considered here,
is injected {\it perpendicular} to the background magnetic field, 
the beam excites electrostatic, standing waves, oscillating at local
plasma frequency, in the beam
injection spatial location.
Here physical situation is different in that now the beam is injected {\it along} the magnetic field which is more plausible
for the type III radio bursts. However, surprisingly we still see in Figure 2(a) the similar effect, that the standing ES
waves are generated in the beam injection location. 
We can estimate the oscillation frequency by counting bright yellow strips in the region
$2 c/\omega_{pe}< x < 8 c/\omega_{pe}$ which is seven starting from the first strip. The time elapsed is $50 \omega_{pe}^{-1}$,
thus $50/2\pi=7.96$ and the conclusion is that this standing wave has approximately the plasma frequency.
Note that the small mismatch is due to the fact that $\omega_{pe}$ in all normalisations
is taken at the edges of the simulation domain where $n_0(0)=n_0(x_{max})=1$, while the beam injection location
is centred on $x_{max}/25$ where $n_0(x_{max}/25)=0.846$. (The beam spatial spread here is within $2 c/\omega_{pe} <x < 8 c/\omega_{pe}$).  
In Figure 2(a) $10 c/\omega_{pe}< x < 35 c/\omega_{pe}$ we see series of oblique strips which
is ES wake created by the beam. This can be evidenced by the fact that slope of these oblique strips is 
$(35-10) (c/\omega_{pe}) / (50 / \omega_{pe}) = 0.5 c$ which coincides with the electron beam injection
speed. Similar conclusion is reached from Figure 2(b) as well where the inferred slope of the beam is the same ($0.5c$).
Note that for the times $t>30 \omega_{pe}^{-1}$ there is small dip formed on the top of the beam (see Figure 2(b)).
However the beam seems to stay intact which would be expected in the quasilinear theory (due to so called beam free streaming).
This is despite the fact that $n_b/n_e = 10^{-3}$ and 
the criterion of weak turbulence regime
of quasilinear theory $\varepsilon \equiv n_b m_e v_b^2 / (n_0 m_e v_{th,e}^2) \ll 1$ \cite{1999SoPh..184..353M} is actually not
met: In our case $\varepsilon = 4.94$ and also the quasilinear relaxation time, $\tau$, 
(time of establishing the plateau in the electron longitudinal velocity distribution function) is given
by $\tau =n_e/({n_b} \omega_{pe})$ (e.g. \cite{1999SoPh..184..353M}). In our cause $\tau =10^3 \omega_{pe}^{-1}$.
Thus, no substantial quasilinear relaxation is expected to take place within $50 \omega_{pe}^{-1}$.
We gather from Figures 2(c)--2(f) as in Section III.a that only low level EM drift wave noise is generated and we see
no generation of regular EM waves, commensurate to type III bursts. 
As discussed in the Introduction section when the beam pitch angle is zero then no EM waves should be generated; 
and it is only for the oblique pitch angles the EM emission generation is possible because then $\vec v_b \cdot \vec E_\perp \not =0$.

\subsection{Inhomogeneous plasma with electron beam injected obliquely  to the magnetic field ($\theta=45^{\circ}$)}
  
\begin{figure*}    
   \centerline{\hspace*{0.015\textwidth}
               \includegraphics[width=0.515\textwidth,clip=]{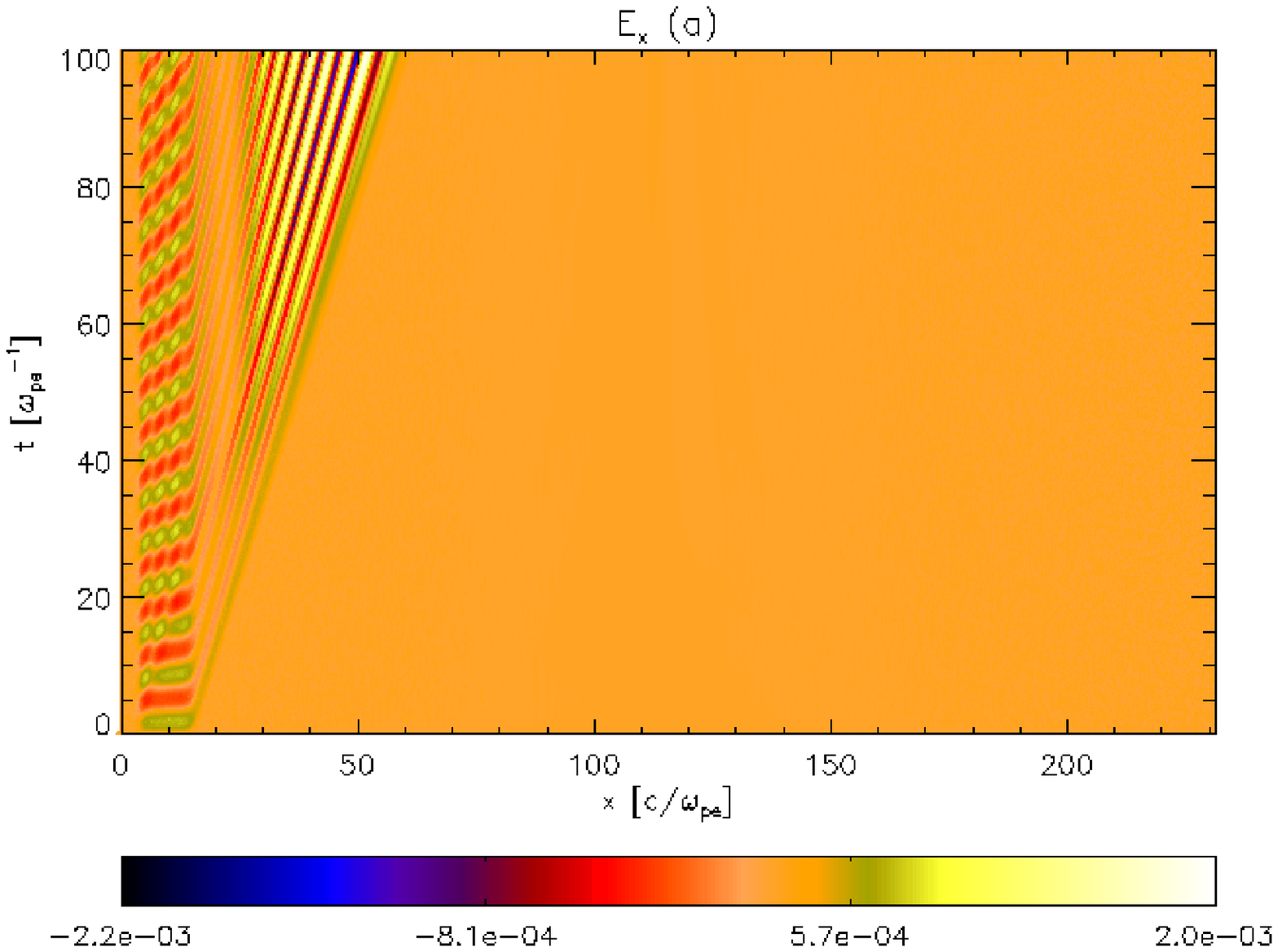}
               \hspace*{-0.03\textwidth}
               \includegraphics[width=0.515\textwidth,clip=]{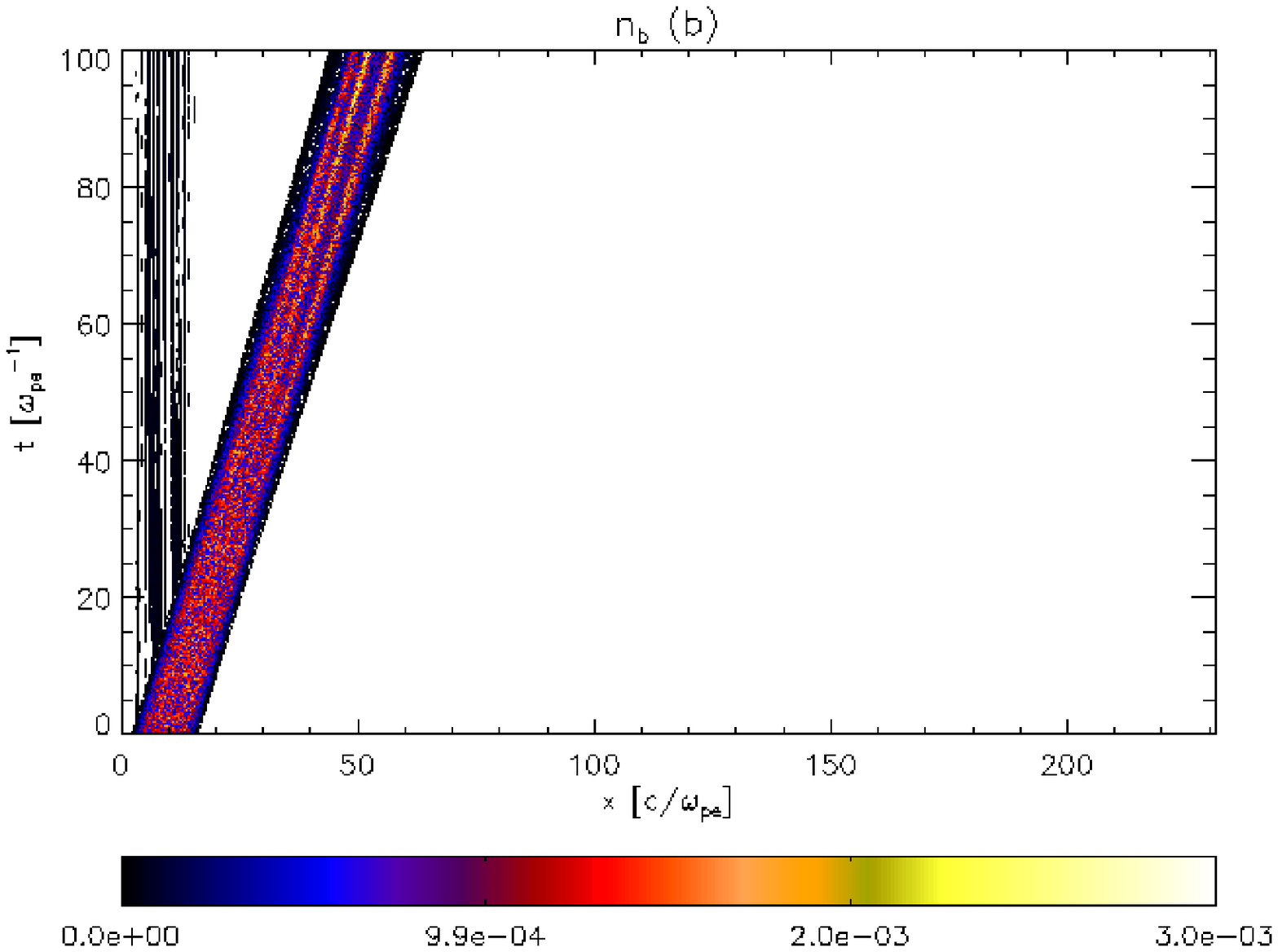}
              }
   \centerline{\hspace*{0.015\textwidth}
               \includegraphics[width=0.515\textwidth,clip=]{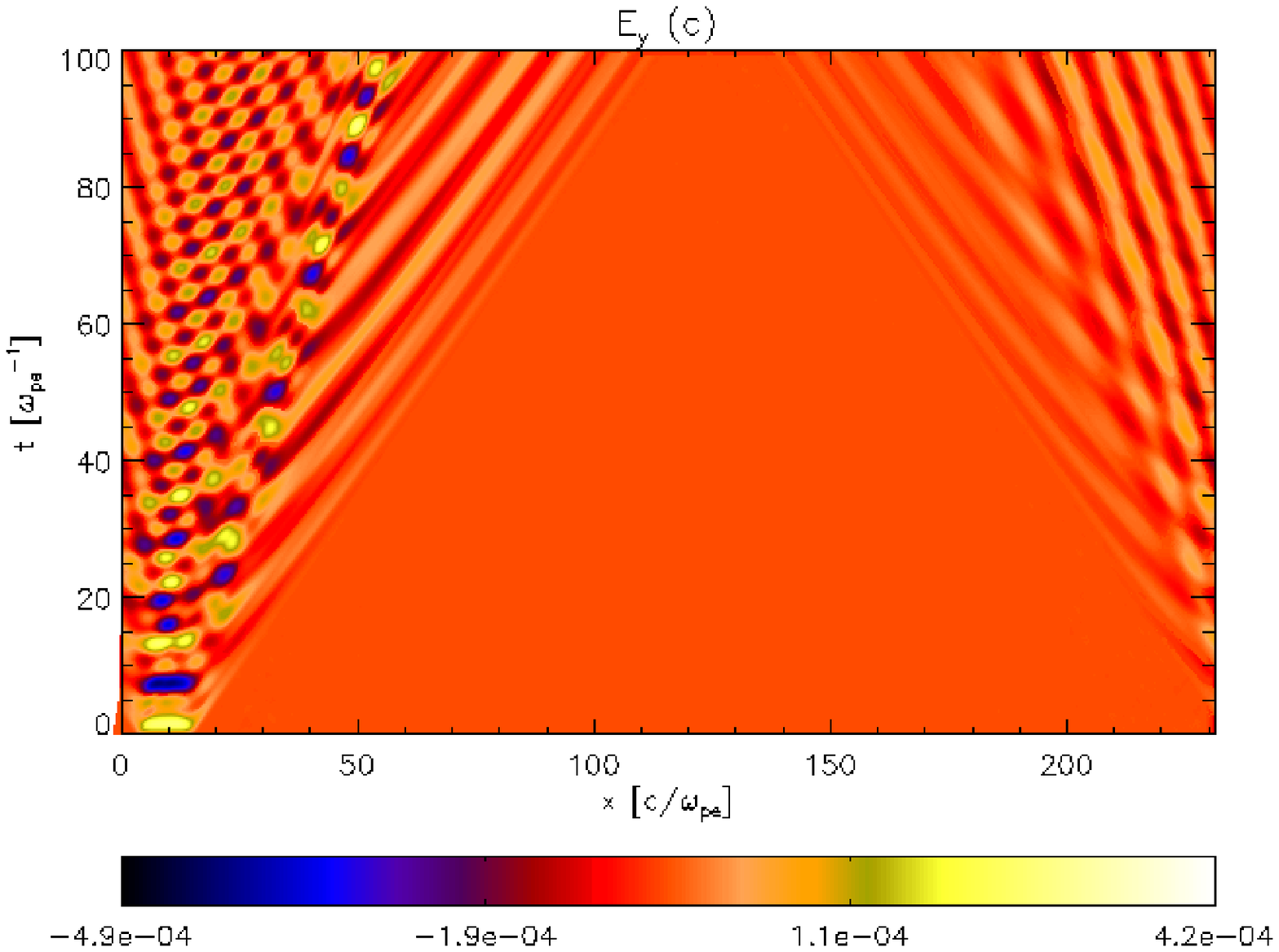}
               \hspace*{-0.03\textwidth}
               \includegraphics[width=0.515\textwidth,clip=]{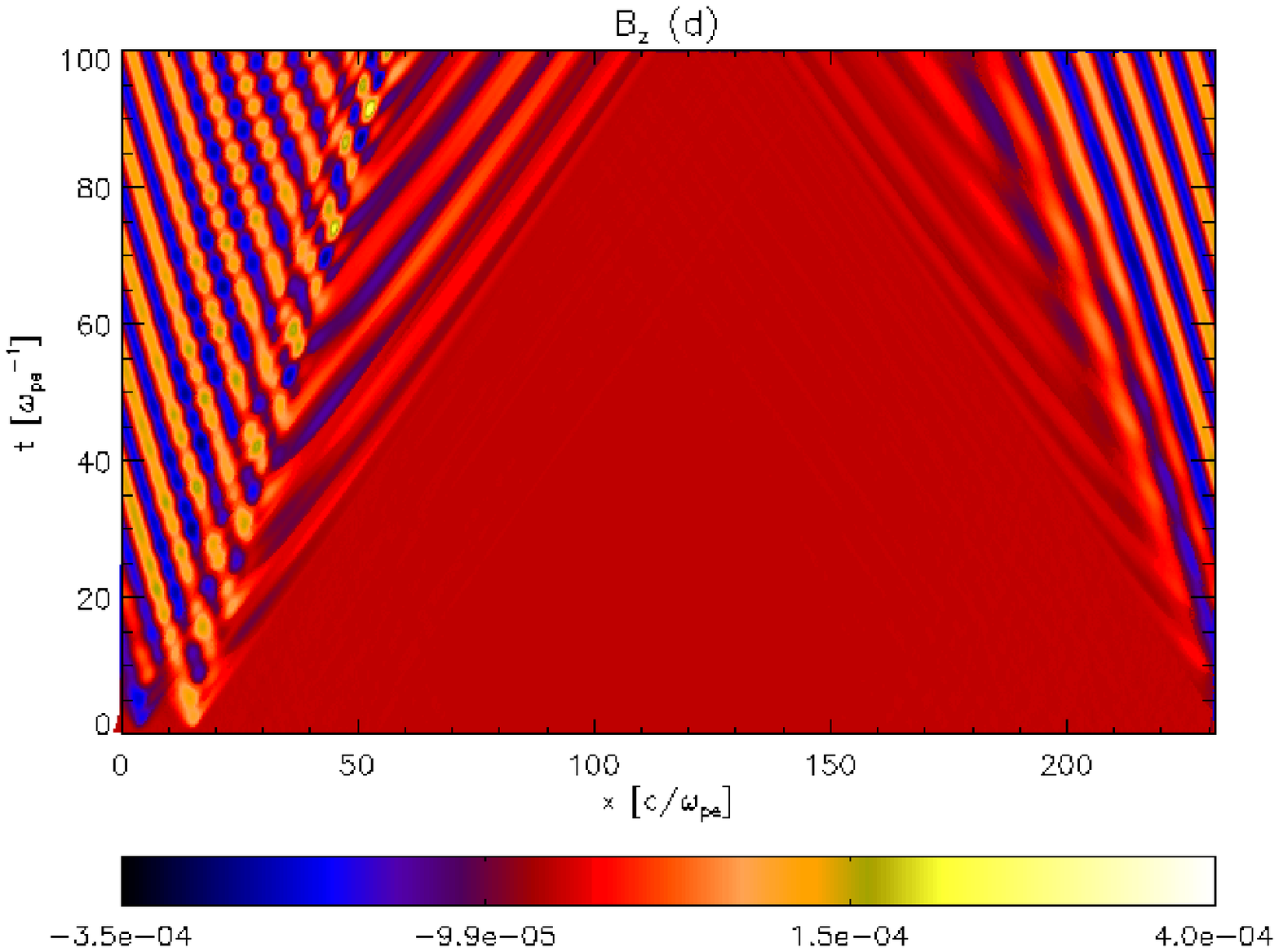}
              }
	      
	      \centerline{\hspace*{0.015\textwidth}
               \includegraphics[width=0.515\textwidth,clip=]{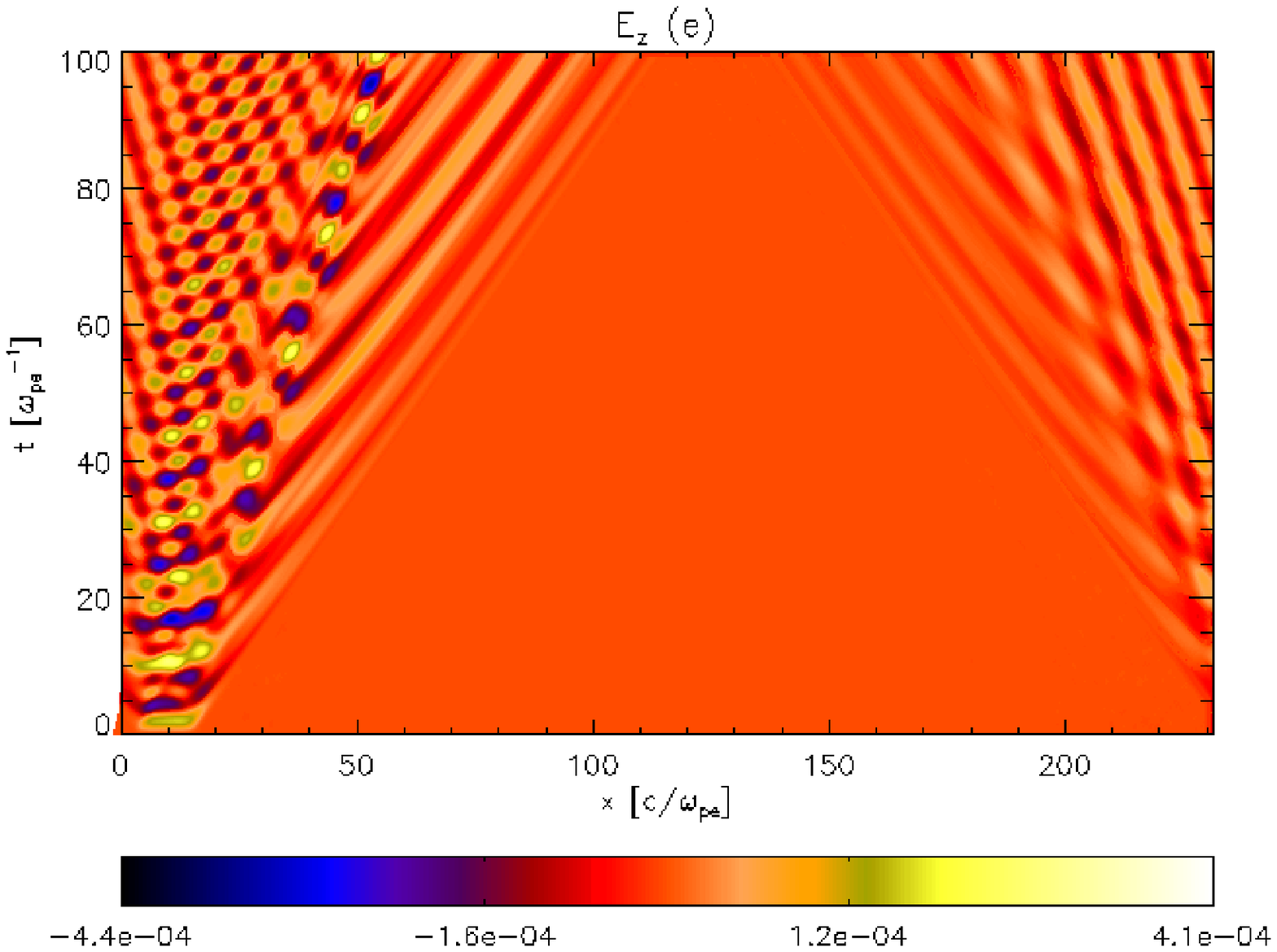}
               \hspace*{-0.03\textwidth}
               \includegraphics[width=0.515\textwidth,clip=]{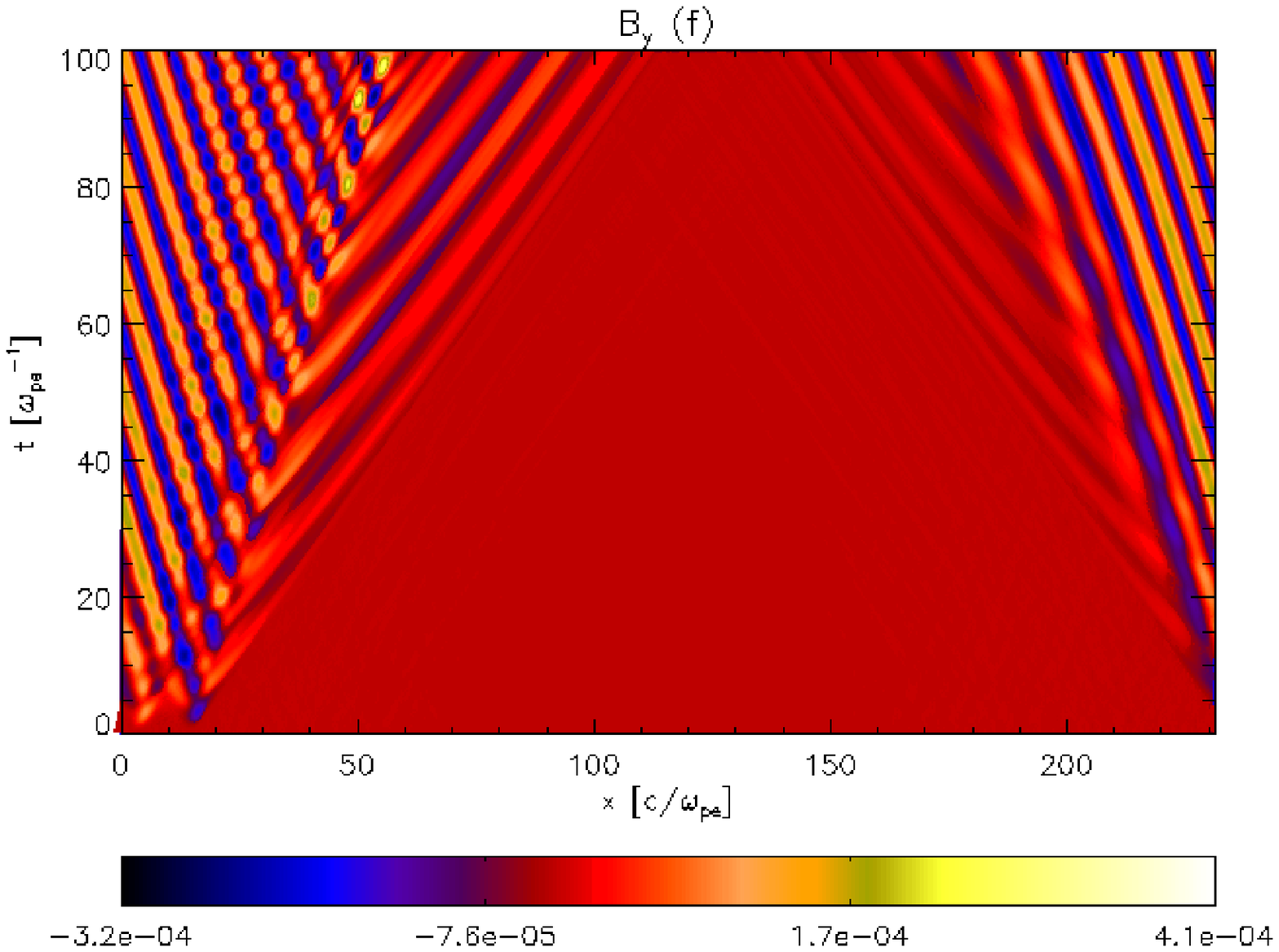}
              }
\caption{As in Figure 2 but for the beam pitch angle $\theta=45^{\circ}$.
This figure pertains to Section III.c.}
   \end{figure*}
In this section we present results when we inject the electron beam  obliquely to the background magnetic field ($\theta =45^{\circ}$).
Here this is achieved by setting $p_{0x}=0.5\gamma m_e c$ and $p_{0y}=0.5\gamma m_e c$ in Eq.(4) at $t=0$.
Note that since we intended to consider a numerical run for a twice longer time ($100 \omega_{pe}^{-1}$) than in other above Sections, 
we have doubled
the spatial domain size whilst keeping the same total number of particles  $1.30065 \times 10^9$.
This implies that now the spatial grid size is $\lambda_D /2$, not $\lambda_D /4$ as in other above Sections.
We gather from Figure 3(a) that again in the beam injection spatial location
standing ES wave oscillating at local plasma frequency is excited. However, the ES wake of the beam,
oblique yellow lines between $30 c/\omega_{pe}< x < 65 c/\omega_{pe}$, detaches from the standing ES wave
and becomes localised. The ES wake also travels with the correct speed of $\approx 0.5c$.
Similar conclusions can be reached from analysing Figure 4(a) which reveals more detailed
spatial structure of the ES wave oscillation and the ES beam wake. Figure 3(b) corroborates the beam travel speed of $0.5c$
as well as reveals minor deviation from the quasilinear theory free streaming, by appearance of
spikes on the top of the beam.
Figures 3(c)--3(f) present time-distance plots for the two components of the transverse electromagnetic fields:
($E_y$ 3(c), $B_z$ 3(d)) and ($E_z$ 3(e), $B_y$ 3(f)).
We learn that as the beam pitch angle now is $\theta=45^{\circ} \not = 0$ escaping EM radiation is generated. 
In the beam injection spatial location,
$4 c/\omega_{pe}< x < 15 c/\omega_{pe}$, we see strong interference pattern between the standing (trapped) ES and escaping
EM radiation. The  EM waves travel in 
both directions, and because of the periodic boundary conditions, waves that travel to the
left, appear on the right side of the simulation domain ($x > 150 c/\omega_{pe}$). 
Figure 4(c) shows normalised electron number density at $t=100 \omega_{pe}^{-1}$
(thick solid curve), beam spatial profiles at $t=0$ (dashed curve) and $t=100 \omega_{pe}^{-1}$
(thin sold curve). Note that when plotting the spatial profiles of 
$n_b$, we scale it by a factor of $10^3$ so that it is clearly visible.
We gather that the background electron population number density stays unchanged throughout the simulation,  compared
to $t=0$. This serves as additional proof that our
initial conditions without the beam are stable. 
Also, comparing Figures 4(a) and 4(c) we confirm that indeed the beam and  ES wake 
travel the same distance at the same speed of $\approx (65-15) c/\omega_{pe}/ (100 / \omega_{pe}) =0.5c$.
Figures 4(b) and 4(d) show a more detailed spatial structure of the transverse, escaping
EM radiation components. We gather that these actually consist of two parts:
(i) the  part within $0< x < 65 c/\omega_{pe}$ corresponds to the non-escaping 
ES standing waves and the
ES wake of the beam; and (ii)  the part within 
$65 c/\omega_{pe} < x < 115 c/\omega_{pe}$ that corresponds to the escaping EM radiation.
Figures 4(e) and 4(f) show electron (including the electron beam) and ion longitudinal velocity distribution function time evolution.
The considered  momentum range is $3 v_{th,i} m_i$, which is then converted to
the velocities (in the relevant Figures) by using $3 v_{th,i} m_i/(\gamma m_{e,i} c)$.
We gather from Figure 4(e) that background electron population distribution remains unchanged
(dashed ($t=0$) and solid ($t=100 \omega_{pe}^{-1}$) curves, centred on $v_x=0$ do overlap to a plotting
accuracy),
while the electron beam starts to show a tendency of plateau formation, according to the
quasi-linear theory. Note that in this Section simulation end time is 100 $\omega_{pe}^{-1}$
while the quasi-linear relaxation time is 1000 $\omega_{pe}^{-1}$. Recall however that we are
not strictly speaking in the quasi-linear regime because $\varepsilon \approx 5$.
The ion distribution function (Figure 4(f)) shows no noticeable by-eye change.
Thus despite the fact that ions in the simulation are treated as mobile, the ion
population shows no dynamics in the velocity space.

Next we attempt to produce a synthetic (simulated) dynamical spectrum. We take a snapshot of 
the spatial profile of one of the transverse EM components, $E_y(x,t=100 \omega_{pe}^{-1})$, and cast it into the
temporal dependence by putting $x= c t$. Thus in Figure 5(a) we see the same pattern
in $E_y$ as in Figure 4(b) but now it appears as function of $t$ normalised to $\omega_{pe}^{-1}$. 
Note that we do not include EM wave which appear on the right due to the periodic boundary
condition, i.e. we restrict ourselves to the range $x=[0,128 c/\omega_{pe}]$ (i.e. the same as 
$t=[0,128 \omega_{pe}^{-1}]$).
We then generate a wavelet power spectrum for the   $E_y(t)$.
Wavelet software was provided by C. Torrence and G. Compo, and is available at URL: 
\url{http://atoc.colorado.edu/research/wavelets/}.
We gather from Figure 5(b) that in the time interval $0< x < 65 \omega_{pe}^{-1}$  
the wavelet power spectrum is flat (period/frequency does not change in time).
This corresponds to ES oscillation part, oscillating at the plasma frequency (with a prefactor of $2\pi$).
In the time interval $65< x < 128 \omega_{pe}^{-1}$  wavelet power spectrum corresponds to
escaping EM radiation part and we clearly see a decrease of EM signal frequency in time.
One has to realise however that in the Figure 5(b) it appears that
large period  $P_{LF}=15/(2\pi)$ (low frequency) at $t \approx 65 \omega_{pe}^{-1}$ shows up first and
then low period $P_{HF}=9/(2\pi)$ (high frequency) at $t \approx 115 \omega_{pe}^{-1}$ follows.
This has a simple explanation that we obtained the time series of $E_y$ by putting  
$x= c t$. In reality for a distant observer located in a point, the high frequency (high density) would appear first followed by
a low frequency radio signal. Note also that the frequency decrease nicely follows the number density decrease.
In  Figure 4(c) dashed line peak which represents the beam at $t=0$ is located at the normalised background electron number 
density
(thick solid line) of 0.846. By the time beam reaches its final destination by time 
$t=100 \omega_{pe}^{-1}$ (thin solid line) the density has dropped to $\approx 0.3$. Therefore we would
have expected that the corresponding plasma frequency has dropped by a factor $\sqrt{0.3/0.846}=0.595$.
Indeed we gather from Figure 5(b) that $P_{HF}/P_{LF} =9/15=0.6$.
In other words, 
$\omega_{p,LF} / \omega_{p,HF} = \sqrt{n_{LF}/n_{HF}}=\sqrt{0.3/0.846}=0.595 \approx P_{HF}/P_{LF}=9/15=0.6$,
where the notation is straightforward.
Therefore we conclude that the frequency decrease in the synthetic dynamical spectrum is commensurate
to the plasma frequency (electron number density) decrease along the beam propagation path.
Note that it is only in the interplanetary type III radio bursts
the frequency drops by many orders of magnitude. However, the bursts that occur
in the solar corona show frequency drops by about a factor of two (as in our simulation), are not
uncommon. For example decameter type II bursts which have a fine structure in the form of 
type III-like bursts. The drift
rates of these sub-bursts are close to the ordinary type III bursts velocity, but their
duration is essentially lesser \cite{2004SoPh..222..151M}.

\begin{figure*}    
   \centerline{\includegraphics[width=0.99\textwidth,clip=]{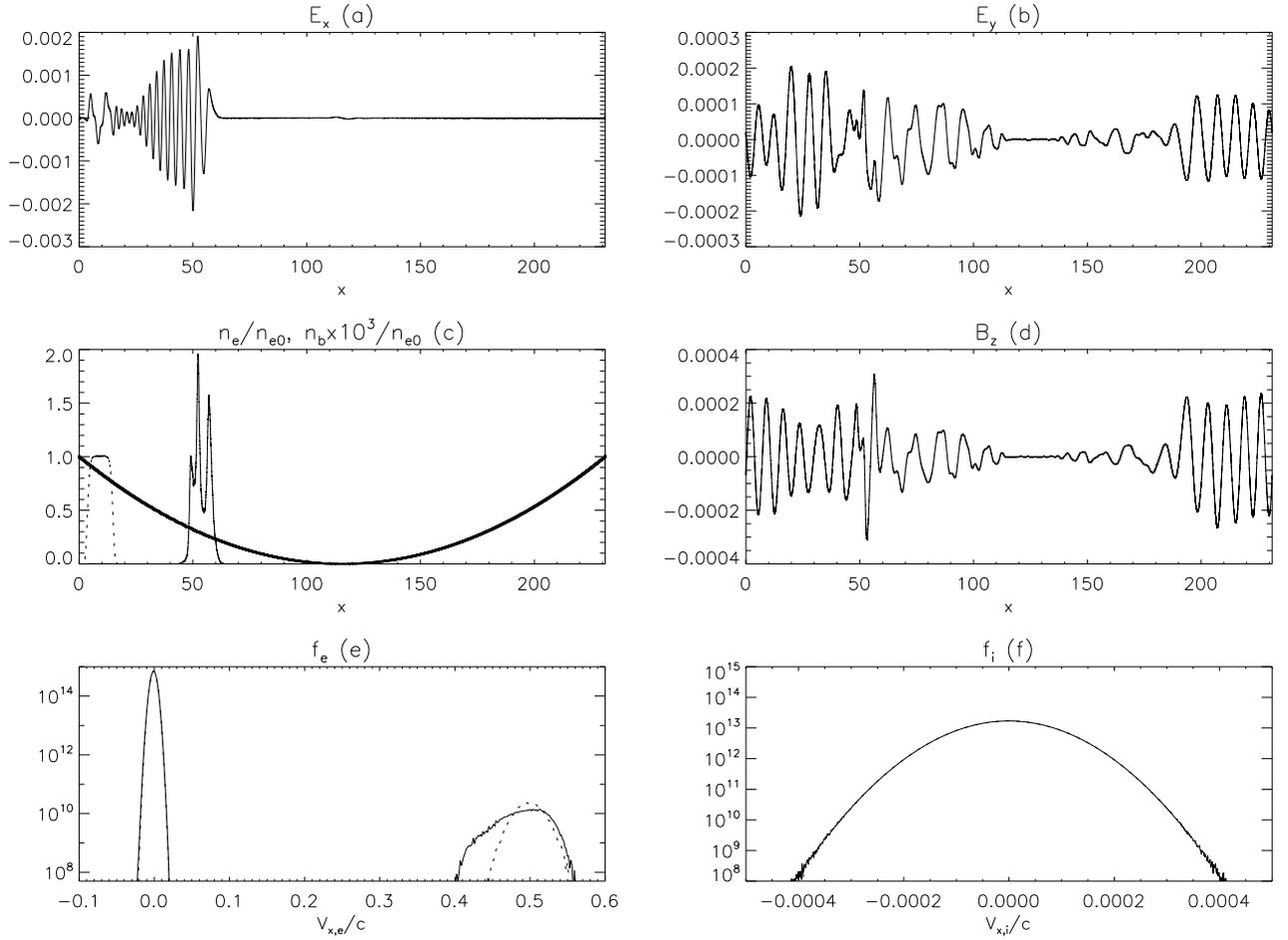}
              }
              \caption{(a) $E_x(x,t=100 \omega_{pe}^{-1})$, (b) $E_y(x,t=100 \omega_{pe}^{-1})$, (c) $n_e(x,t=100 \omega_{pe}^{-1})$ (thick solid curve), 
$n_b(x,t=100 \omega_{pe}^{-1})$ (thin solid curve), $n_b(x,t=0)$ (dashed curve)
(Note that $n_b$ is scaled by a factor of $10^3$ so that it is clearly visible.), (d) $B_z(x,t=100 \omega_{pe}^{-1})$,
(e) background and beam electron longitudinal ($v_x$) velocity distribution functions $f_e(v_x,t=100 \omega_{pe}^{-1})
+f_b(v_x,t=100 \omega_{pe}^{-1})$ (solid curve)
and $f_e(v_x,t=0)+f_b(v_x,t=0)$ (dashed curve), (f) ion longitudinal ($v_x$) velocity distribution functions $f_i(v_x,t=100 \omega_{pe}^{-1})$ (solid curve)
and $f_i(v_x,t=0)$ (dashed curve) (Note that to a plotting precision the two curves overlap). This figure pertains to Section III.c.}
      \end{figure*}

\begin{figure}    
   \centerline{\includegraphics[width=0.5\textwidth,clip=]{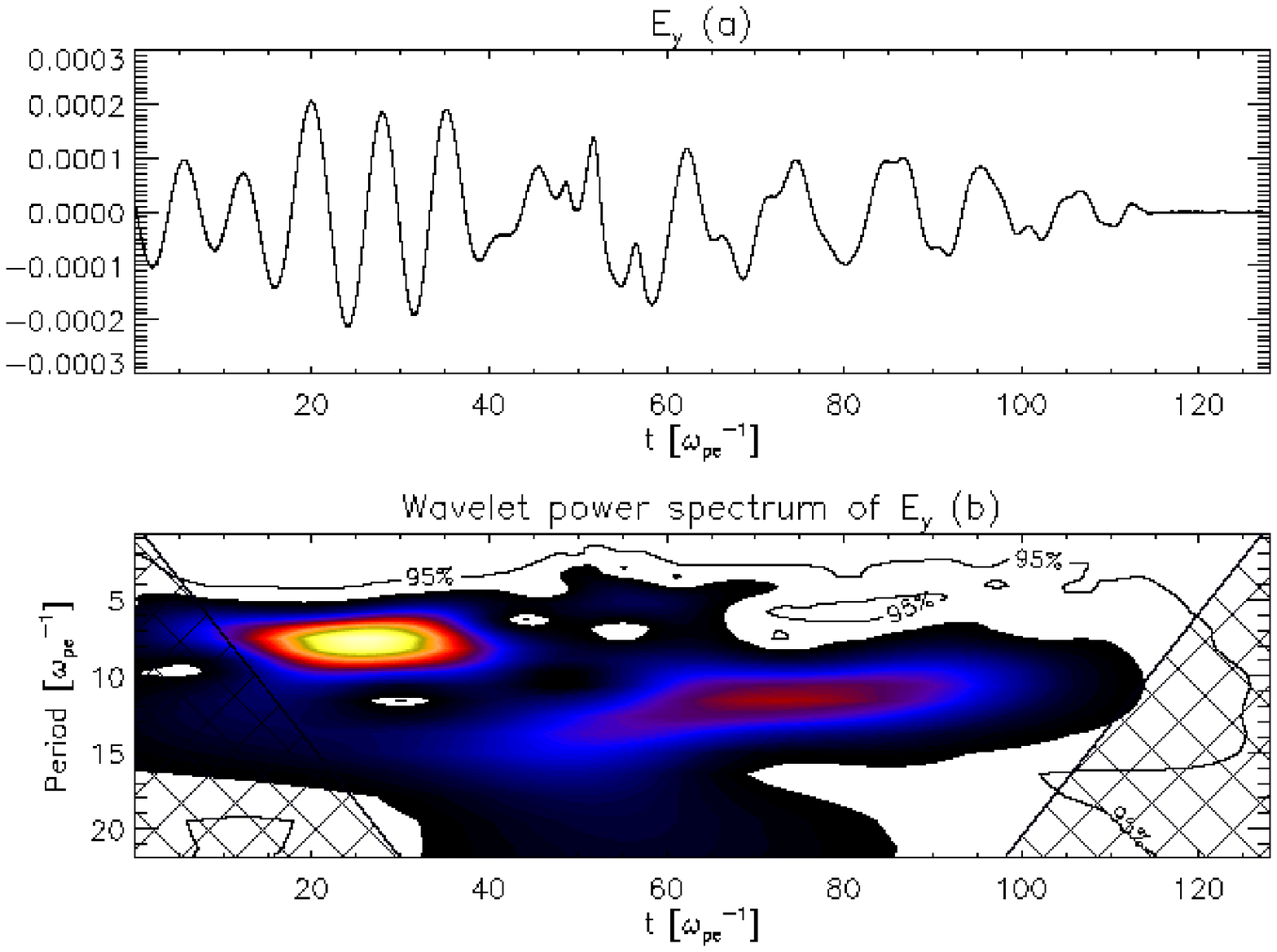}
              }
              \caption{(a) Time evolution of $E_y$ (see text how spatial $x$-dependence was cast into time $t$-dependence), 
	      (b) wavelet power spectrum of (a). This figure pertains to Section III.c.}
      \end{figure}

\subsection{Homogeneous plasma with electron beam injected obliquely to the magnetic field ($\theta=45^{\circ}$)}
\begin{figure*}    
   \centerline{\hspace*{0.015\textwidth}
               \includegraphics[width=0.515\textwidth,clip=]{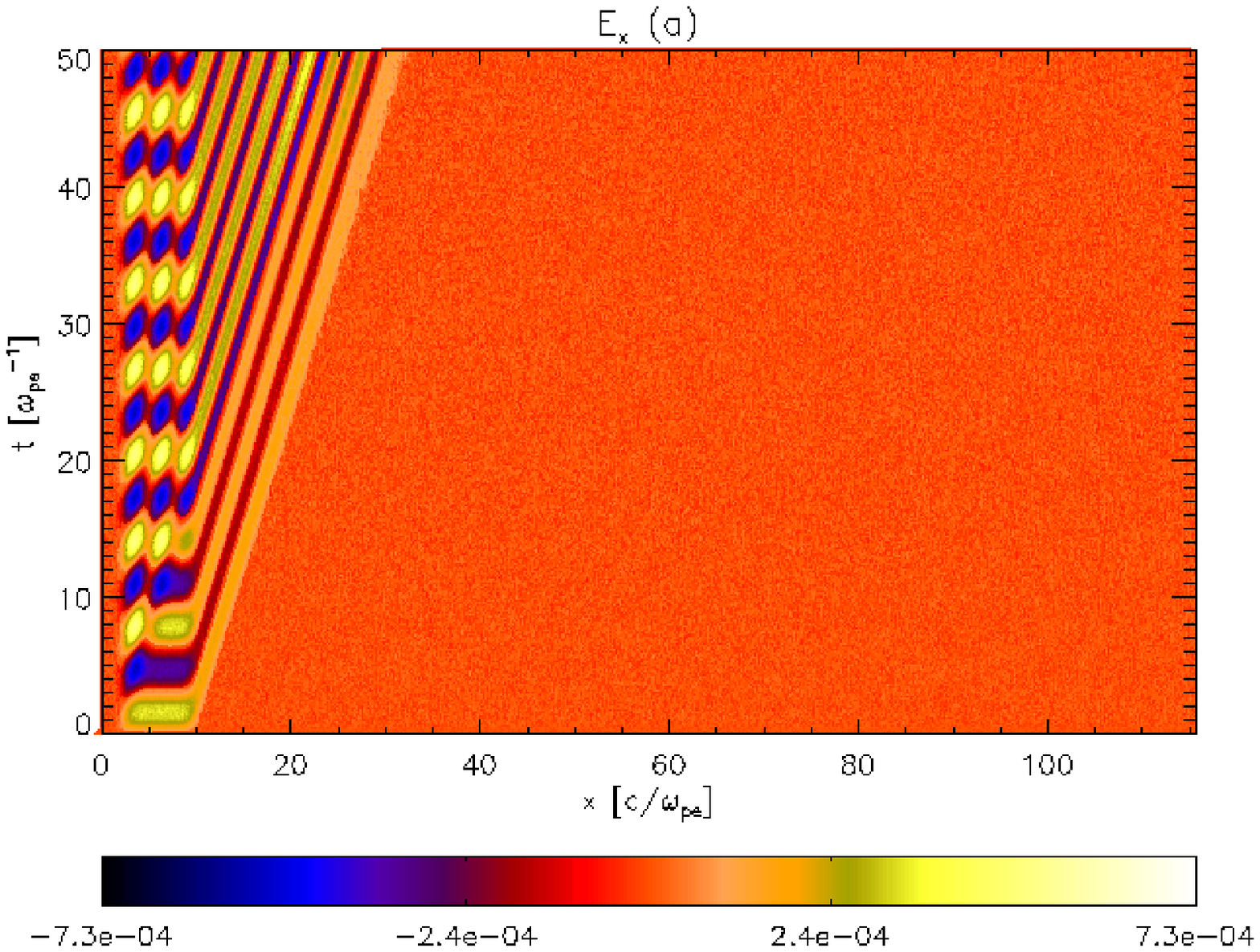}
               \hspace*{-0.03\textwidth}
               \includegraphics[width=0.515\textwidth,clip=]{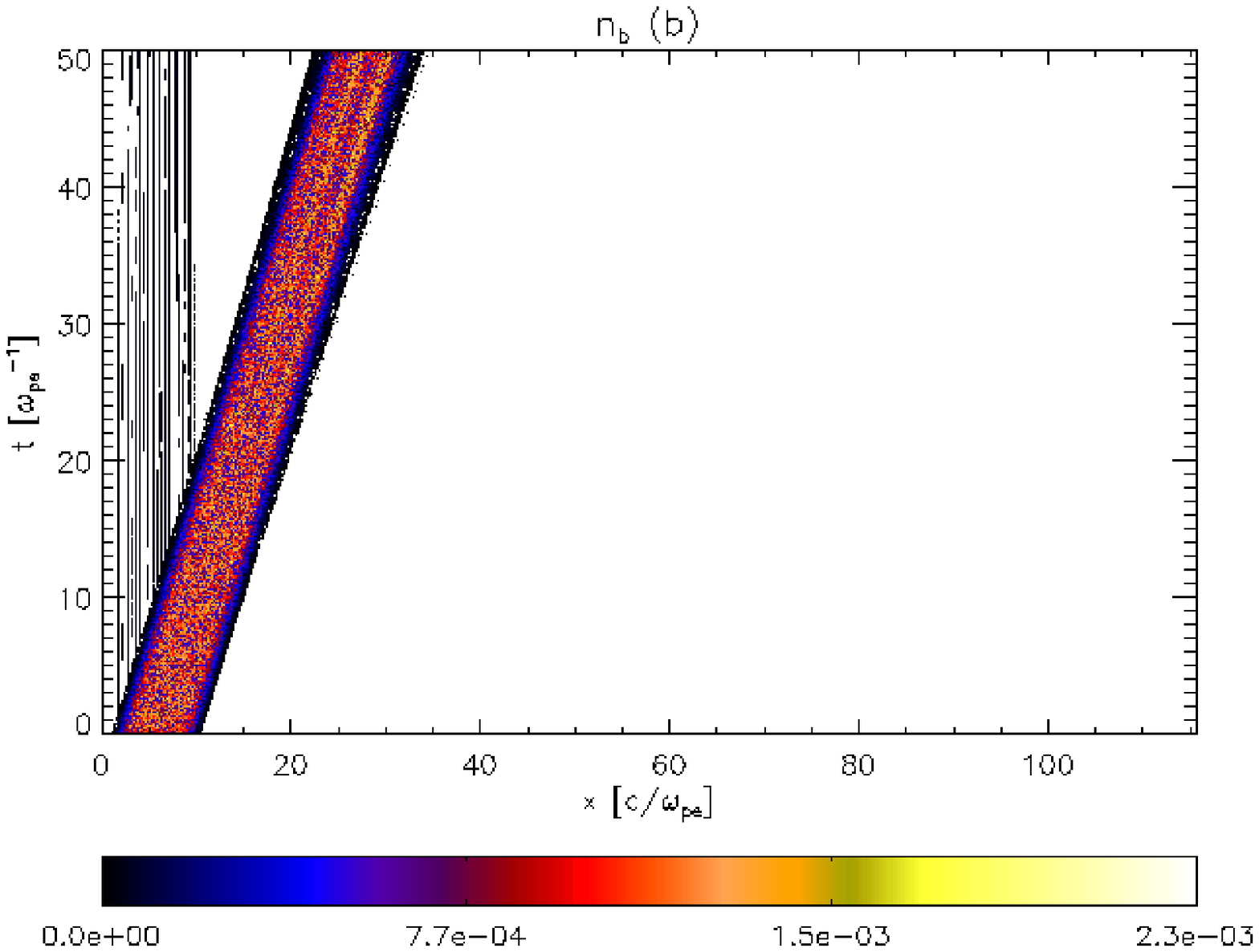}
              }
   \centerline{\hspace*{0.015\textwidth}
               \includegraphics[width=0.515\textwidth,clip=]{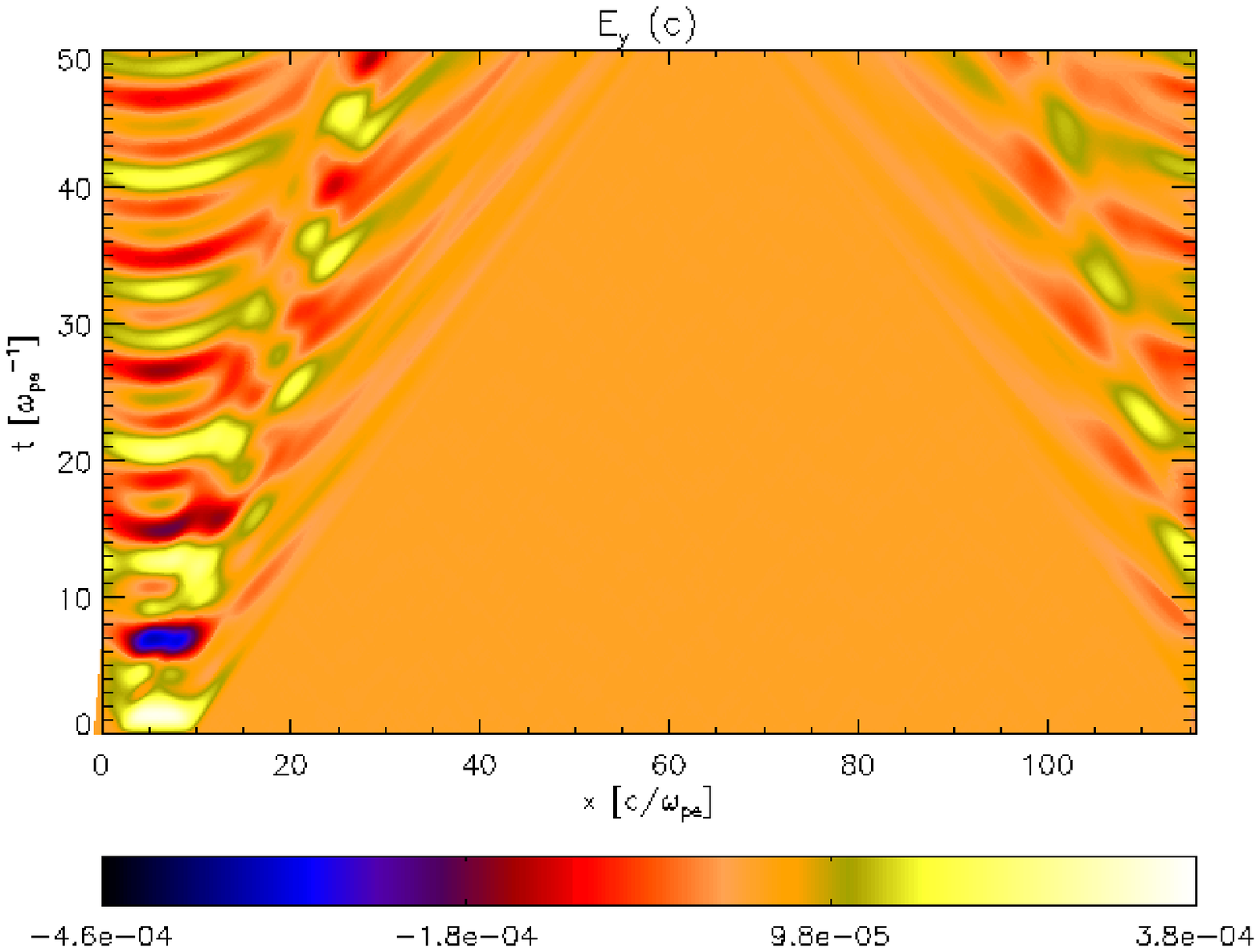}
               \hspace*{-0.03\textwidth}
               \includegraphics[width=0.515\textwidth,clip=]{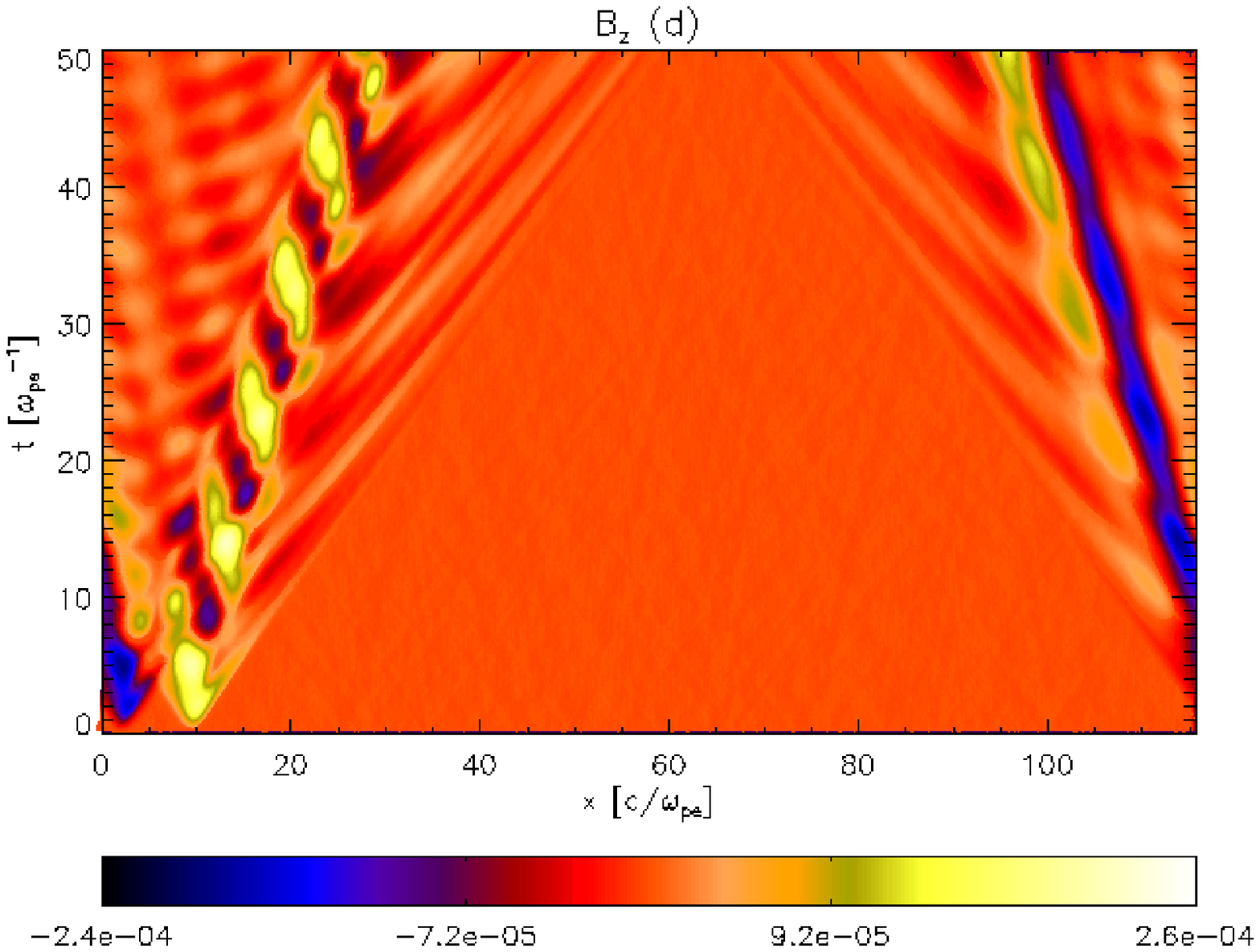}
              }
	      
	      \centerline{\hspace*{0.015\textwidth}
               \includegraphics[width=0.515\textwidth,clip=]{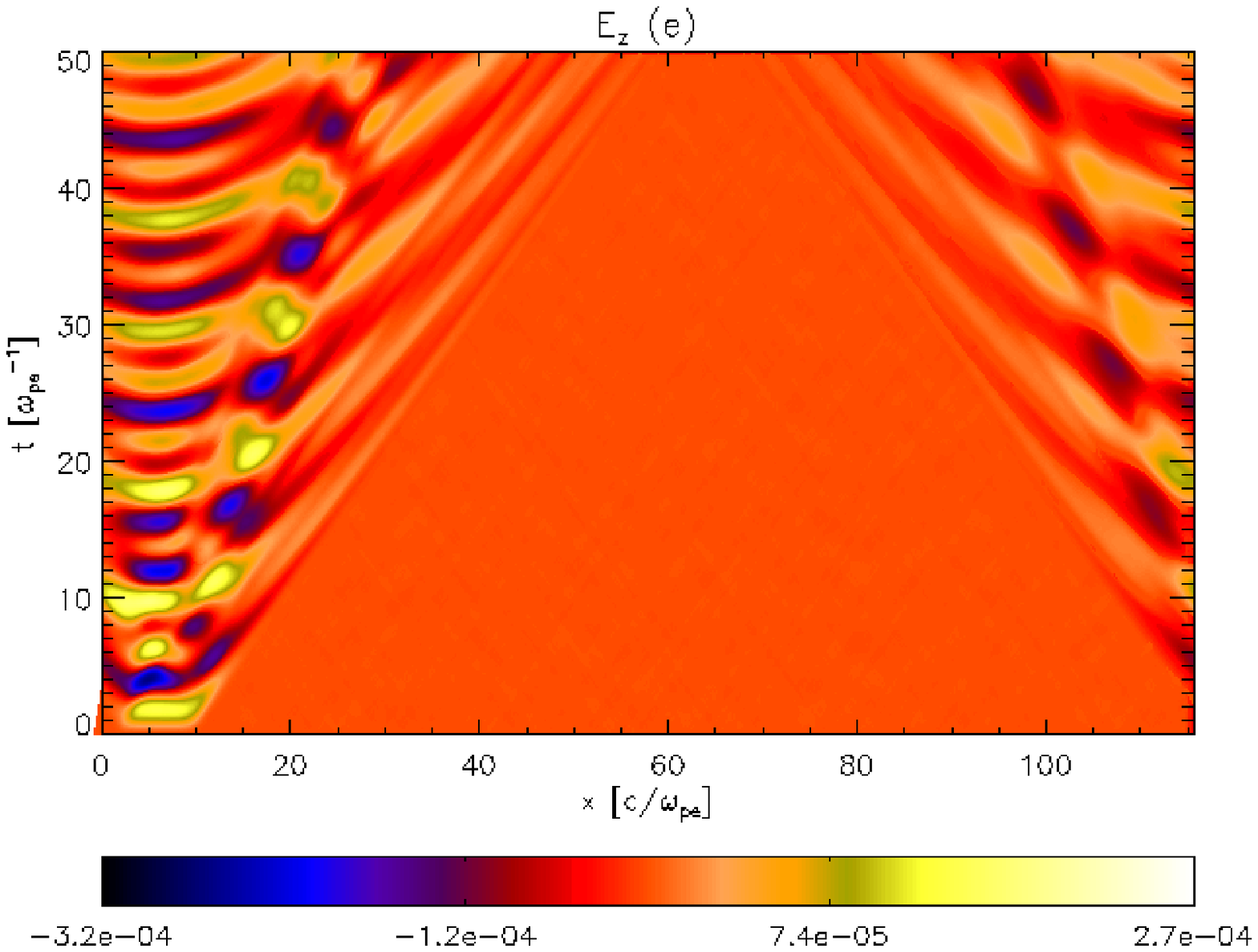}
               \hspace*{-0.03\textwidth}
               \includegraphics[width=0.515\textwidth,clip=]{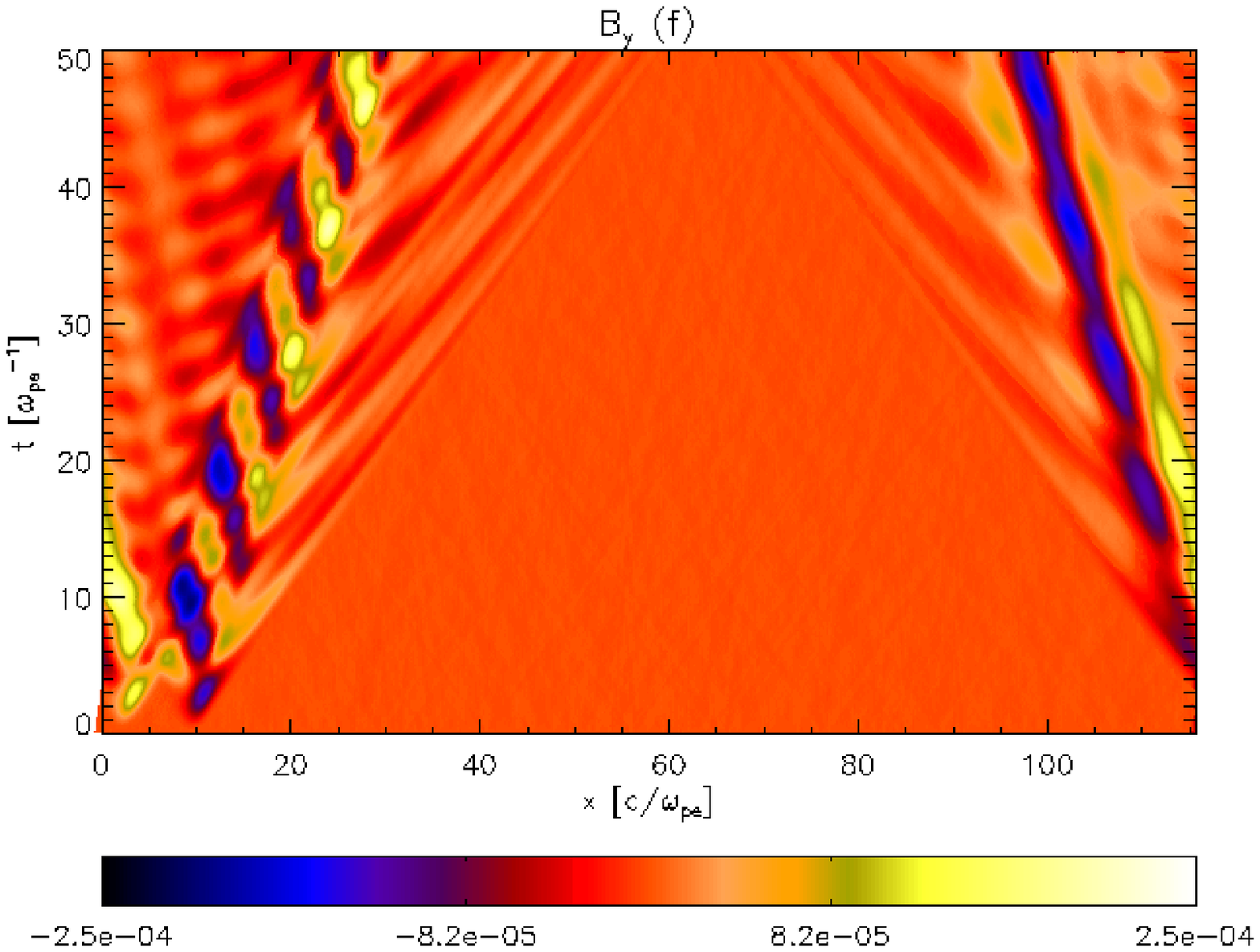}
              }
\caption{As in Figure 2 but for the beam pitch angle $\theta=45^{\circ}$ and homogeneous background number density.
This figure pertains to Section III.d.}
   \end{figure*}      
   
\begin{figure*}    
   \centerline{\includegraphics[width=0.99\textwidth,clip=]{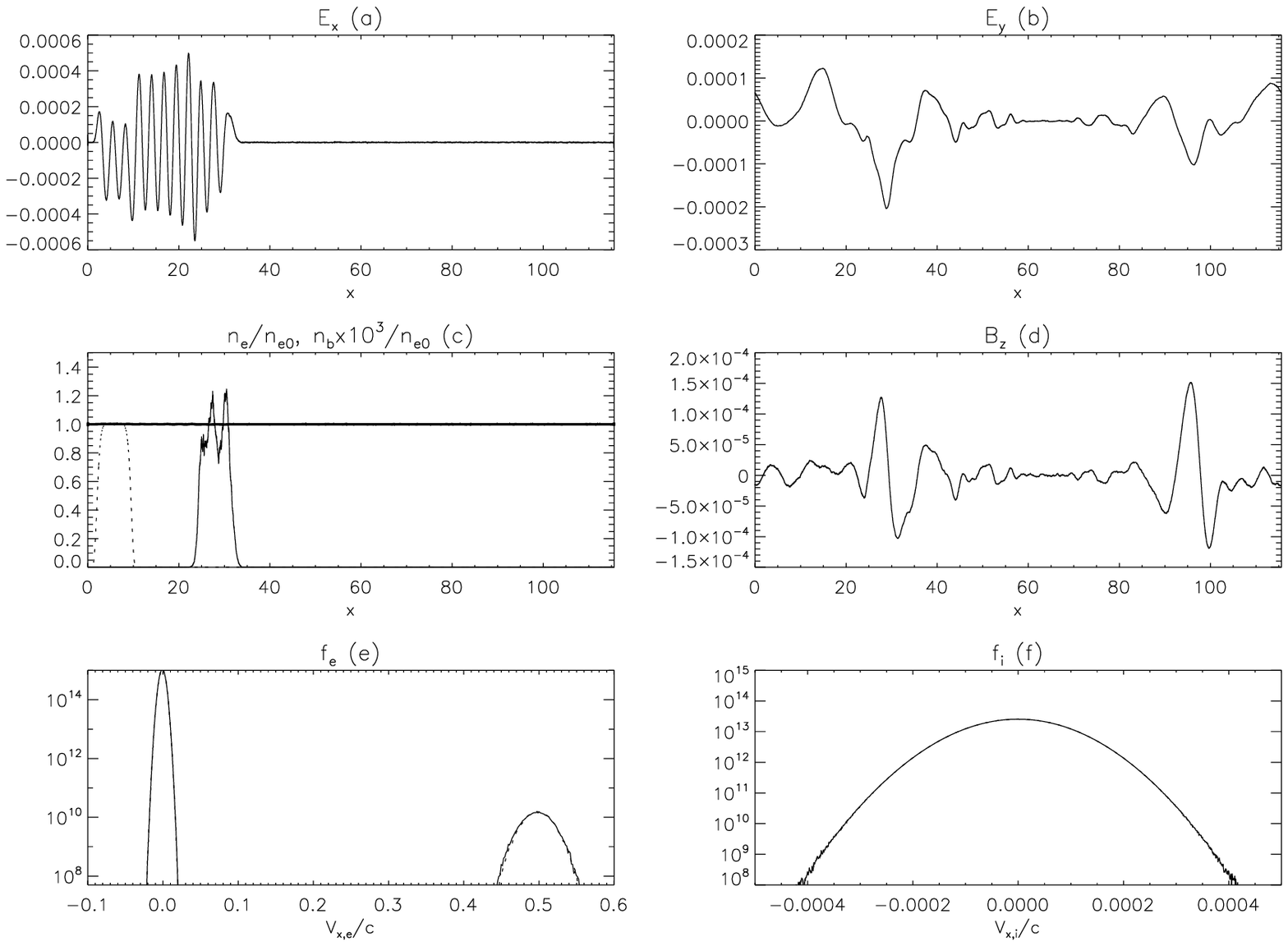}
              }
              \caption{(a) $E_x(x,t=50 \omega_{pe}^{-1})$, (b) $E_y(x,t=50 \omega_{pe}^{-1})$, (c) $n_e(x,t=50 \omega_{pe}^{-1})$ (thick solid line), 
$n_b(x,t=50 \omega_{pe}^{-1})$ (thin solid curve), $n_b(x,t=0)$ (dashed curve) 
(Note that $n_b$ is scaled by a factor of $10^3$ so that it is clearly visible.), (d) $B_z(x,t=50 \omega_{pe}^{-1})$,
(e) background and beam electron longitudinal ($v_x$) velocity distribution functions $f_e(v_x,t=50 \omega_{pe}^{-1})+
f_b(v_x,t=50 \omega_{pe}^{-1})$ (solid curve)
and $f_e(v_x,t=0)+f_b(v_x,t=0)$ (dashed curve), (f) ion longitudinal ($v_x$) velocity distribution functions $f_i(v_x,t=50 \omega_{pe}^{-1})$ (solid curve)
and $f_i(v_x,t=0)$ (dashed curve) (Note that to a plotting precision the two curves overlap). This figure pertains to Section III.d. }
      \end{figure*}

In this section we present the results when we inject the electron beam  obliquely to the background magnetic field ($\theta =45^{\circ}$)
as in Section III.c.
However, contrary to the Section III.c here we consider a plasma with uniform background number density $n_0=1$.
We see from Figure 6(a) that as in Section III.c in the location where the beam was injected standing
ES waves are generated, oscillating at local plasma frequency. However, as better seen from Figure 7(a) now the
ES wake created by the beam does not have enough time to detach itself from the standing wave. This is due to the fact that this
is a shorter run ($t_{end}=50\omega_{pe}^{-1}$). We also gather from Figures 6(b) and 7(c) that the beam travels the correct distance,
commensurate to its speed.
Figures 6(c)--6(f) show time distance plots of transverse EM components that are generated by the Langmuir (ES) waves.
Their more detailed spatial profiles are shown in Figures 7(b) and 7(d).
It is evident by eye that in the uniform plasma case there is no frequency decrease with time for the generated EM components.
Therefore we do not produce the synthetic dynamical spectrum as in Section III.c.
Figures 7(e) confirms that the predictions of the quasilinear theory, that by time $t=50\omega_{pe}^{-1}$
we see (i) no noticeable quasi-linear relaxation because the plateau is expected to develop in 
quasi-linear relaxation time of 1000 $\omega_{pe}^{-1}$; 
(ii) electron free streaming is also evident.
No noticeable change in ion velocity 
distribution function is seen either.
Note that the escaping EM radiation is generated in the uniform plasma density case too. This indicates that
the density gradient plays no role in the EM emission generation.

\subsection{Inhomogeneous plasma with electron beam injected obliquely  to the magnetic field ($\theta=45^{\circ}$),
weak magnetic field case}

In this Section we consider the case similar to III.c but with ten times weaker magnetic field, $B=3$ gauss,
such that $\omega_{ce}/\omega_{pe}(x=0)=0.094$, unlike in the rest of the paper where
$\omega_{ce}/\omega_{pe}(x=0)=0.94$ (which is more appropriate to solar coronal conditions). 
This is to unambiguously demonstrate that
the present study is indeed relevant for the type III radio burst emission,
via the 1.5D non-zero pitch angle (non-gyrotropic) electron beam
quasilinear relaxation, and subsequent emission at the plasma frequency rather than the electron 
gyro-frequency emission.
In the laboratory plasma there are microwave generation devices, such as Gyrotron,
in which EM radiation is produced at electron gyro-frequency, $\omega_{ce}$.
At first sight, it would seem probable that since for the considered model 
parameters throughout this paper $\omega_{ce}/\omega_{pe}(x=0)=0.94$, what we report is
the Gyrotron type EM radiation. The issue can be settled by e.g. lowering the magnetic field
value. The results are presented in Figs. 8-10, which are mirror analogs to
Figs.3-5, except
that here  $\omega_{ce}/\omega_{pe}(x=0)=0.094$.
We gather from Fig.8a that the generated ES (Langmuir) component behaviour is
nearly identical to that of Fig.3a, i.e. we again see the generation of
standing ES waves,  oscillating at plasma frequency, $\omega_{pe}$,
in the beam injection spatial location,
$4 c/\omega_{pe}< x < 15 c/\omega_{pe}$. 
The electron beam dynamics is also identical
(cf. Fig.8b and Fig.3b). There are notable differences in the generated transverse to the
magnetic field EM components: By comparing Figs.8c--8f to  Figs.3c--3f,
we see that in the escaping EM ration there is no interference pattern,
i.e. there is no interference between the standing (trapped) ES and escaping
EM radiation. This is because now ES oscillation no longer
appears in $E_y$, $E_z$, $B_y$, and $B_z$ transverse EM components.
What is {\it crucial} that in the time-distance plots, Figs.8c--8f, we still 
see the {\it same} number of bright (and dark) strips, which is
indicative of the fact that the escaping EM radiation oscillates again
at approximately  {\it plasma frequency}, $\omega_{pe}$, and not
at electron cyclotron frequency $\omega_{ce}$.

\begin{figure*}    
   \centerline{\hspace*{0.015\textwidth}
               \includegraphics[width=0.515\textwidth,clip=]{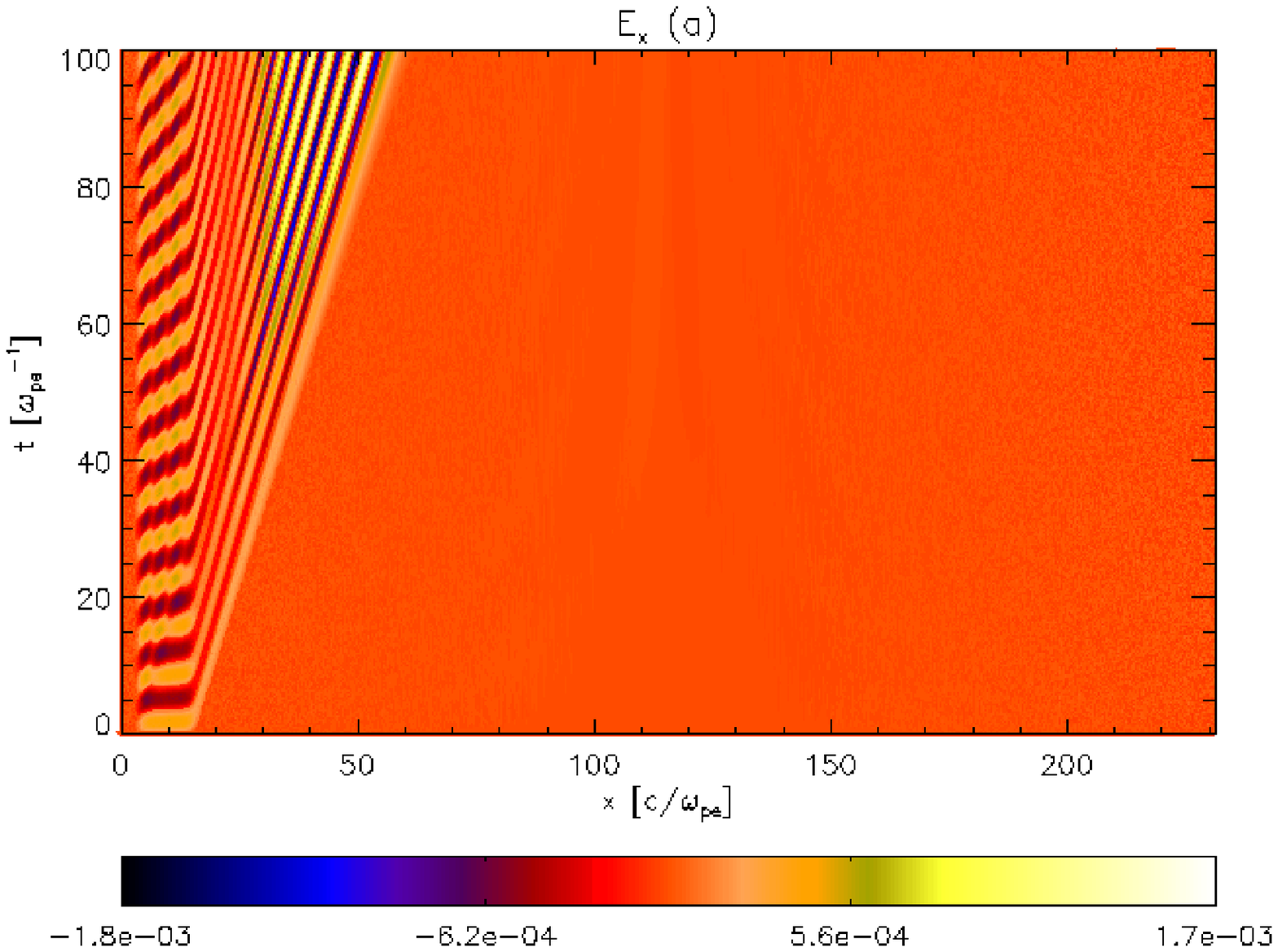}
               \hspace*{-0.03\textwidth}
               \includegraphics[width=0.515\textwidth,clip=]{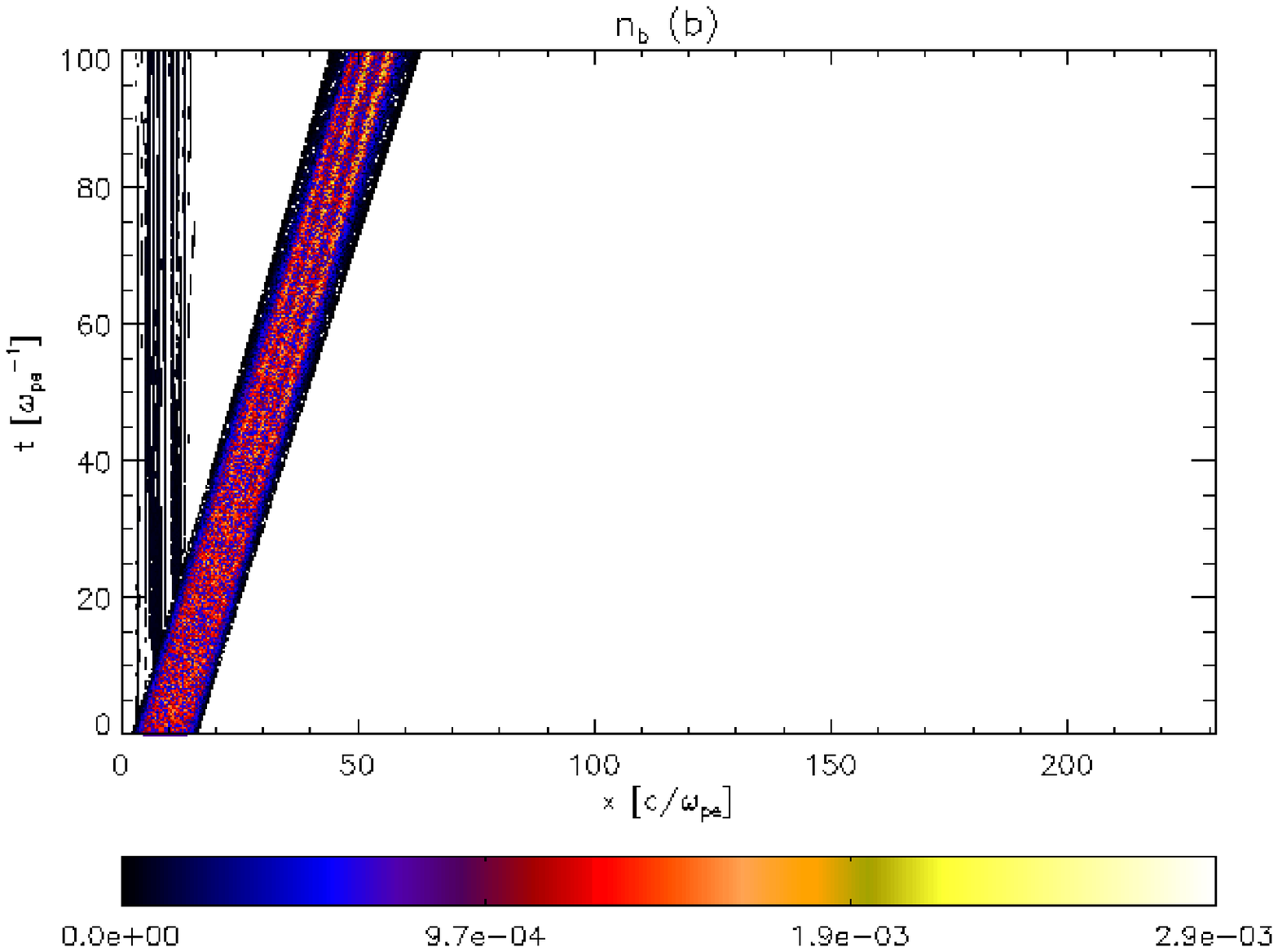}
              }
   \centerline{\hspace*{0.015\textwidth}
               \includegraphics[width=0.515\textwidth,clip=]{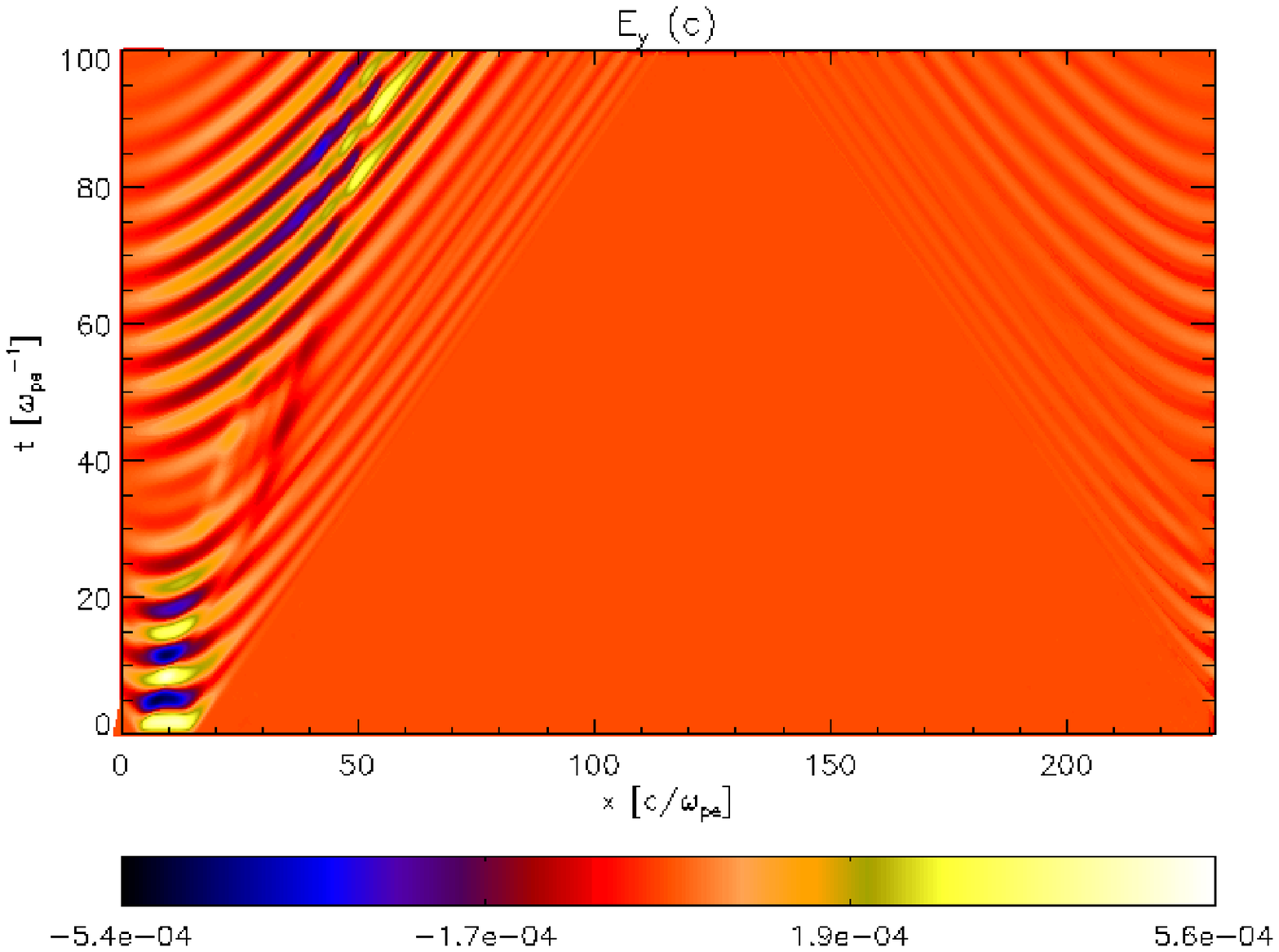}
               \hspace*{-0.03\textwidth}
               \includegraphics[width=0.515\textwidth,clip=]{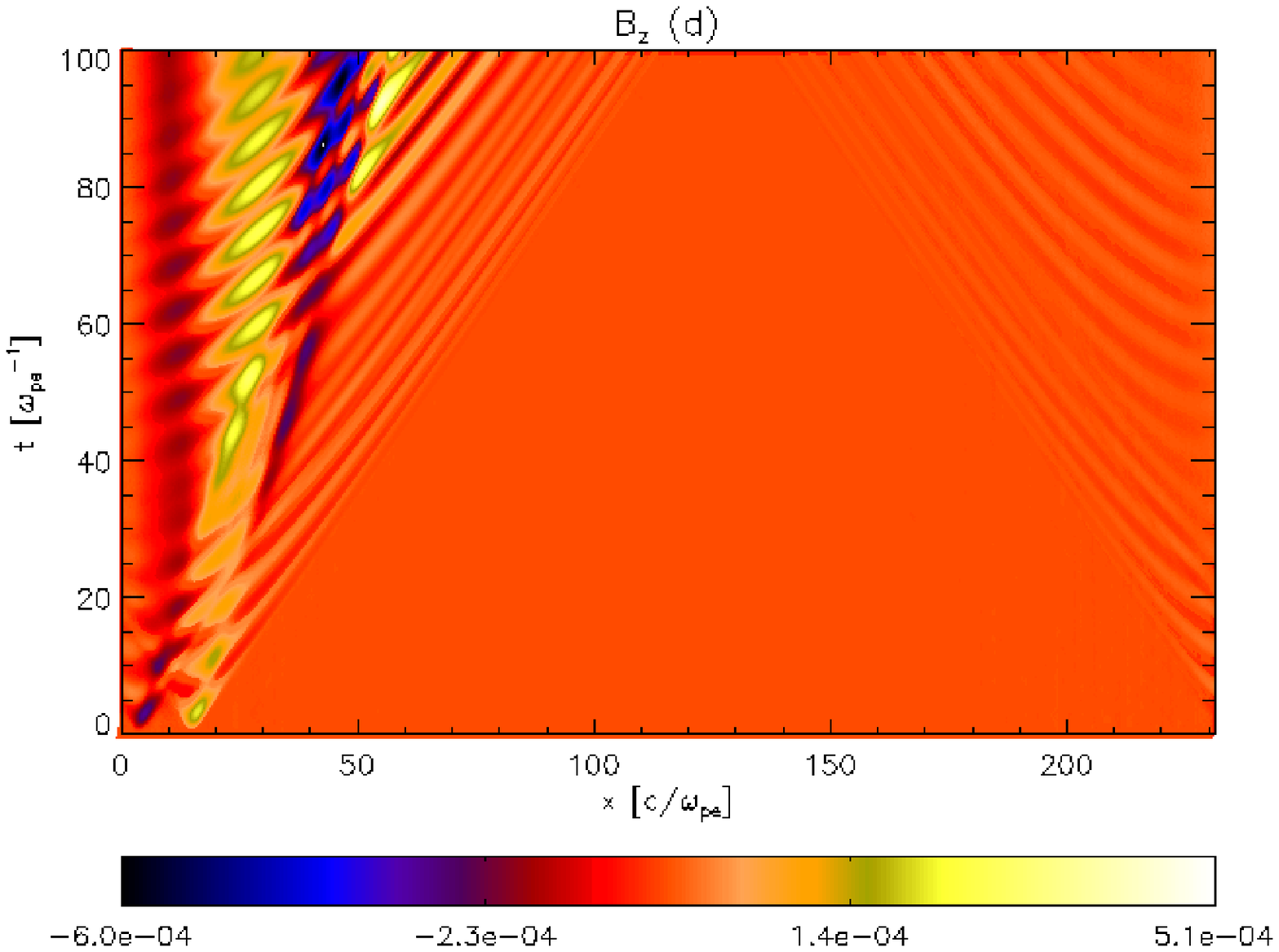}
              }
	      
	      \centerline{\hspace*{0.015\textwidth}
               \includegraphics[width=0.515\textwidth,clip=]{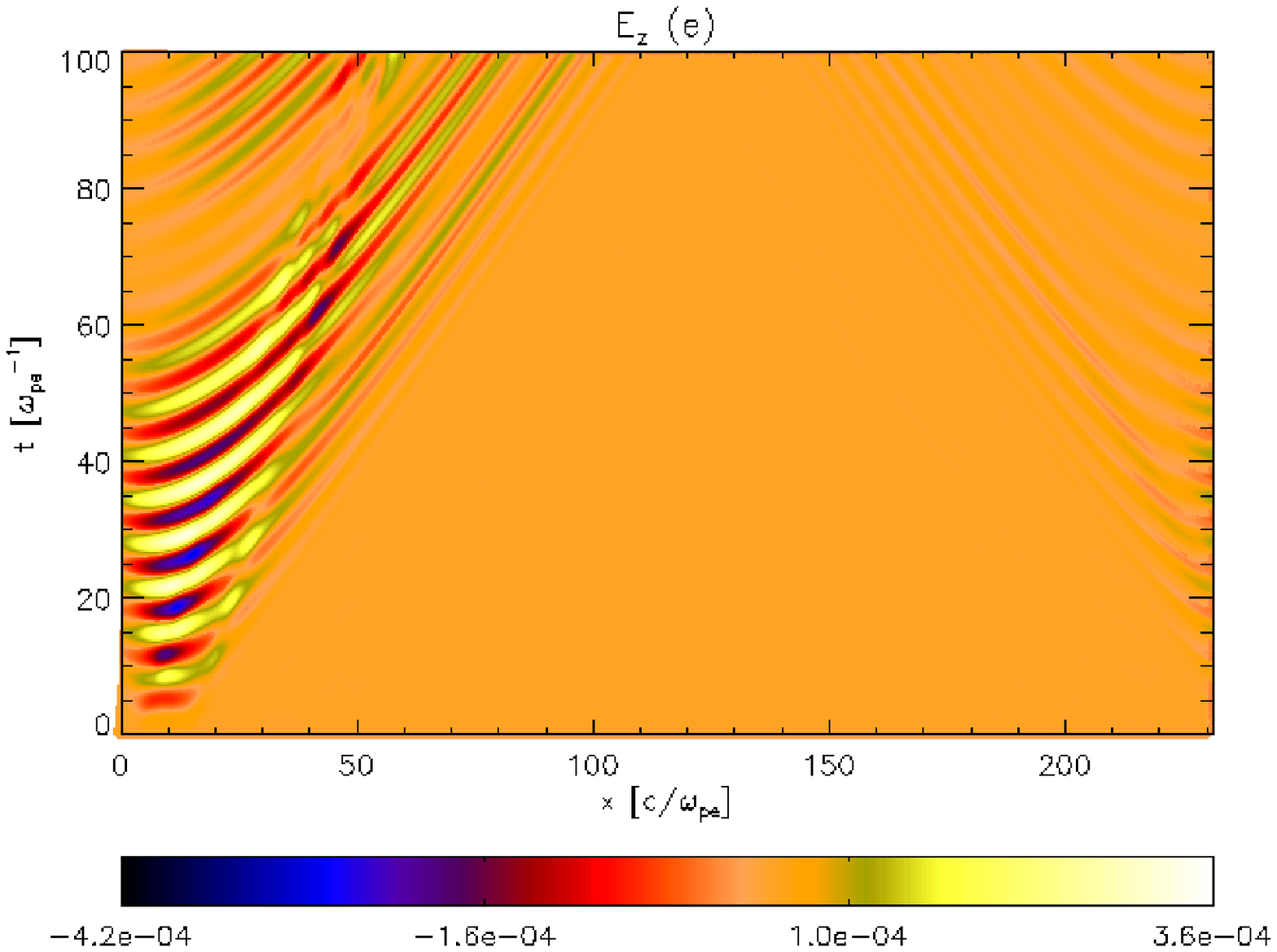}
               \hspace*{-0.03\textwidth}
               \includegraphics[width=0.515\textwidth,clip=]{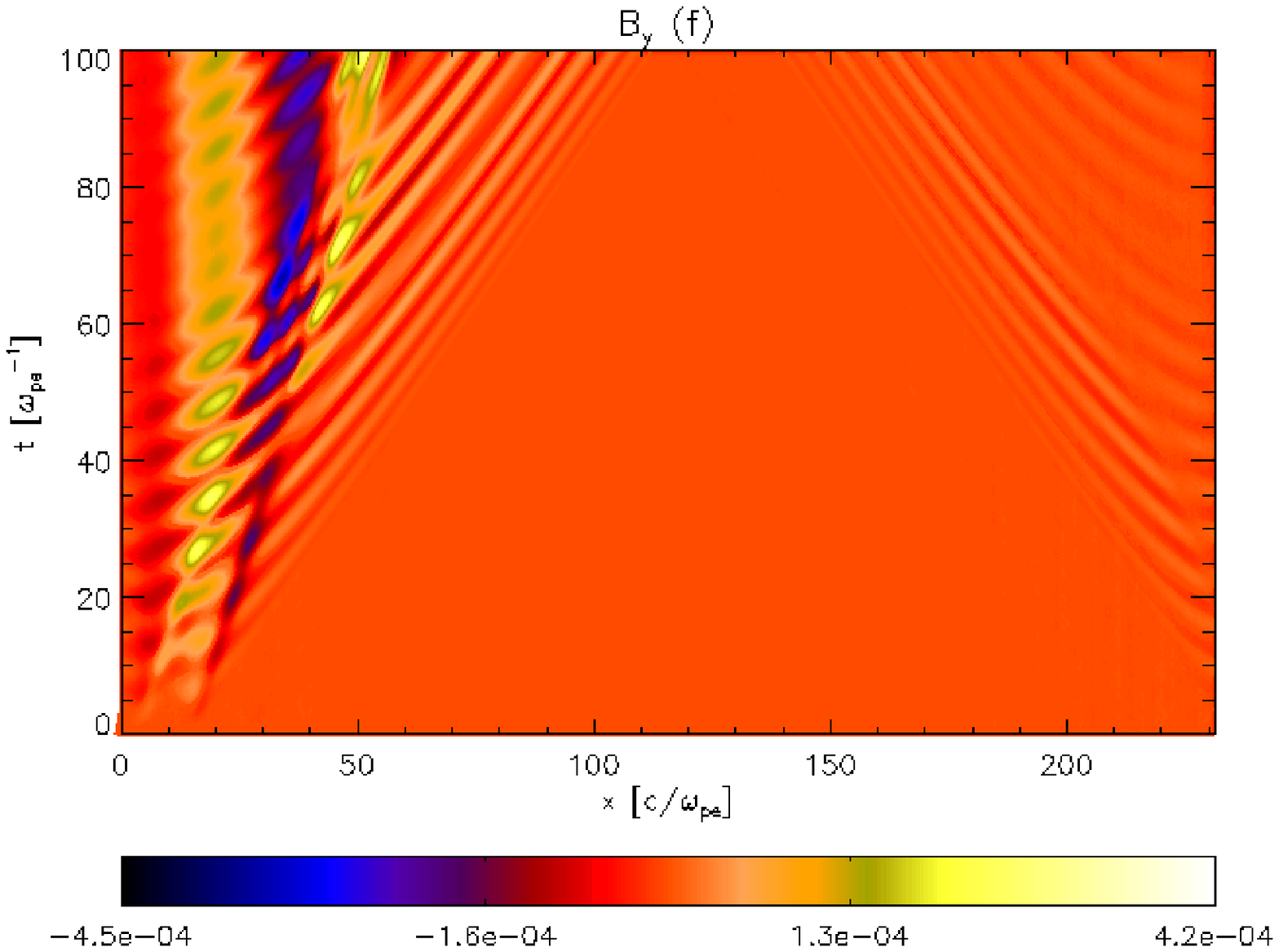}
              }
\caption{As in Figure 3 but for 10 times weaker magnetic field such that $\omega_{ce}/\omega_{pe}(x=0)=0.094$.
This figure pertains to Section III.e.}
   \end{figure*}   

We gather from Fig.9, which corresponds closely to Fig.4, but here $\omega_{ce}/\omega_{pe}(x=0)=0.094$,
that most of the conclusions reached when considering Fig. 4 apply here too.
However, a notable difference is that by comparing Figs.4b and 9b we deduce that
in transverse electric field component $E_y$, for $0< x < 65 c/\omega_{pe}$,
 we no longer see the standing (trapped)
ES component, i.e. only the escaping EM radiation is present.

\begin{figure*}    
   \centerline{\includegraphics[width=0.99\textwidth,clip=]{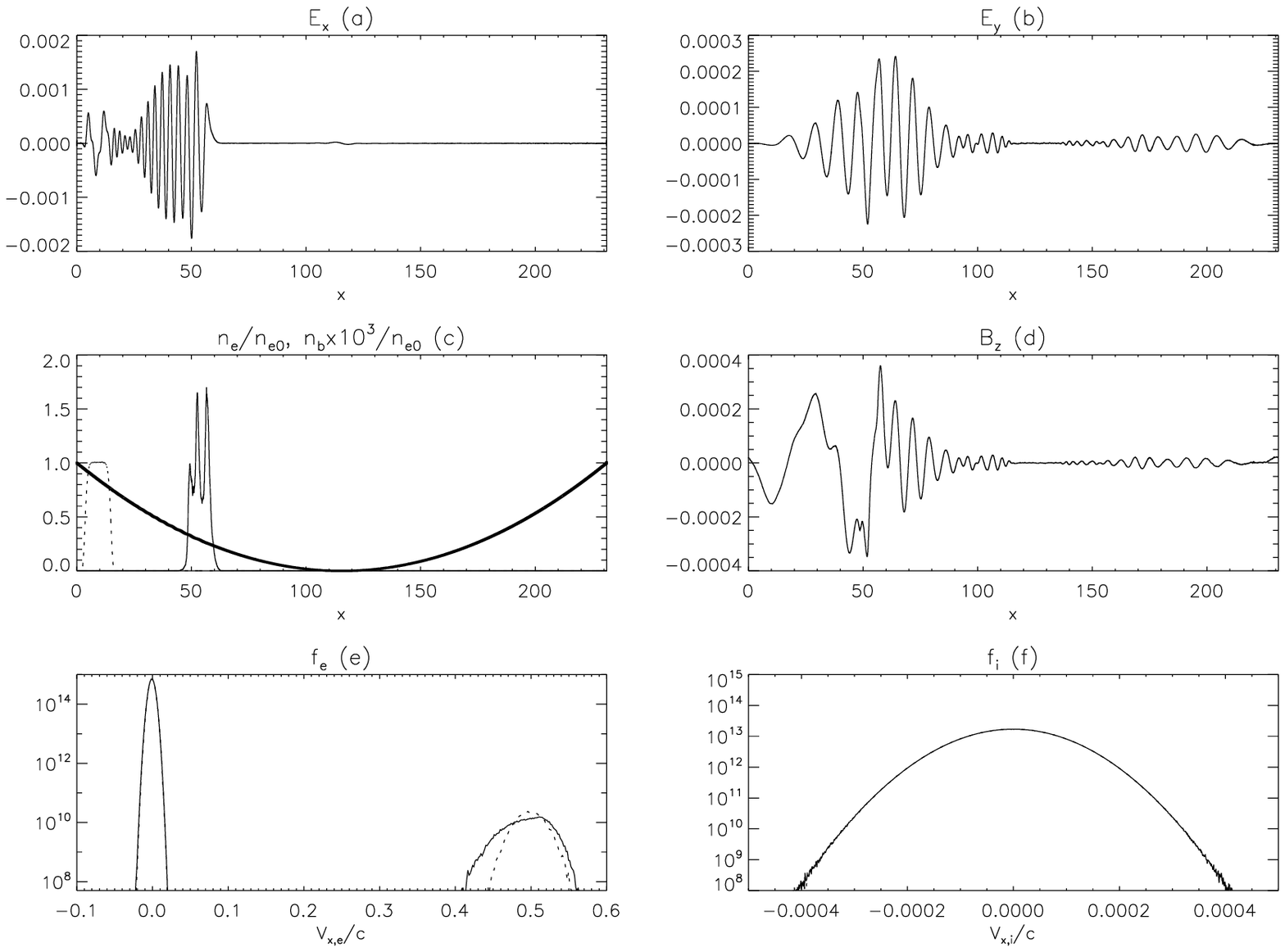}
              }
              \caption{As in Fig.4 but for 10 times weaker magnetic field such that $\omega_{ce}/\omega_{pe}(x=0)=0.094$.
	      This figure pertains to Section III.e.}
      \end{figure*}

Fig.10 has been produced in the same way as Fig.5, except that now $\omega_{ce}/\omega_{pe}(x=0)=0.094$.
We gather from Fig.10 that, contrary to Fig.5 in which both the ES oscillation and escaping EM component
were present, now we see only escaping EM wave which clearly shows the drift towards lower
frequencies. We also observe that as in Fig.5, $\omega_{p,LF} / \omega_{p,HF} \approx 0.6$ that is commensurate
to the background plasma density decrease.
   
\begin{figure}    
   \centerline{\includegraphics[width=0.5\textwidth,clip=]{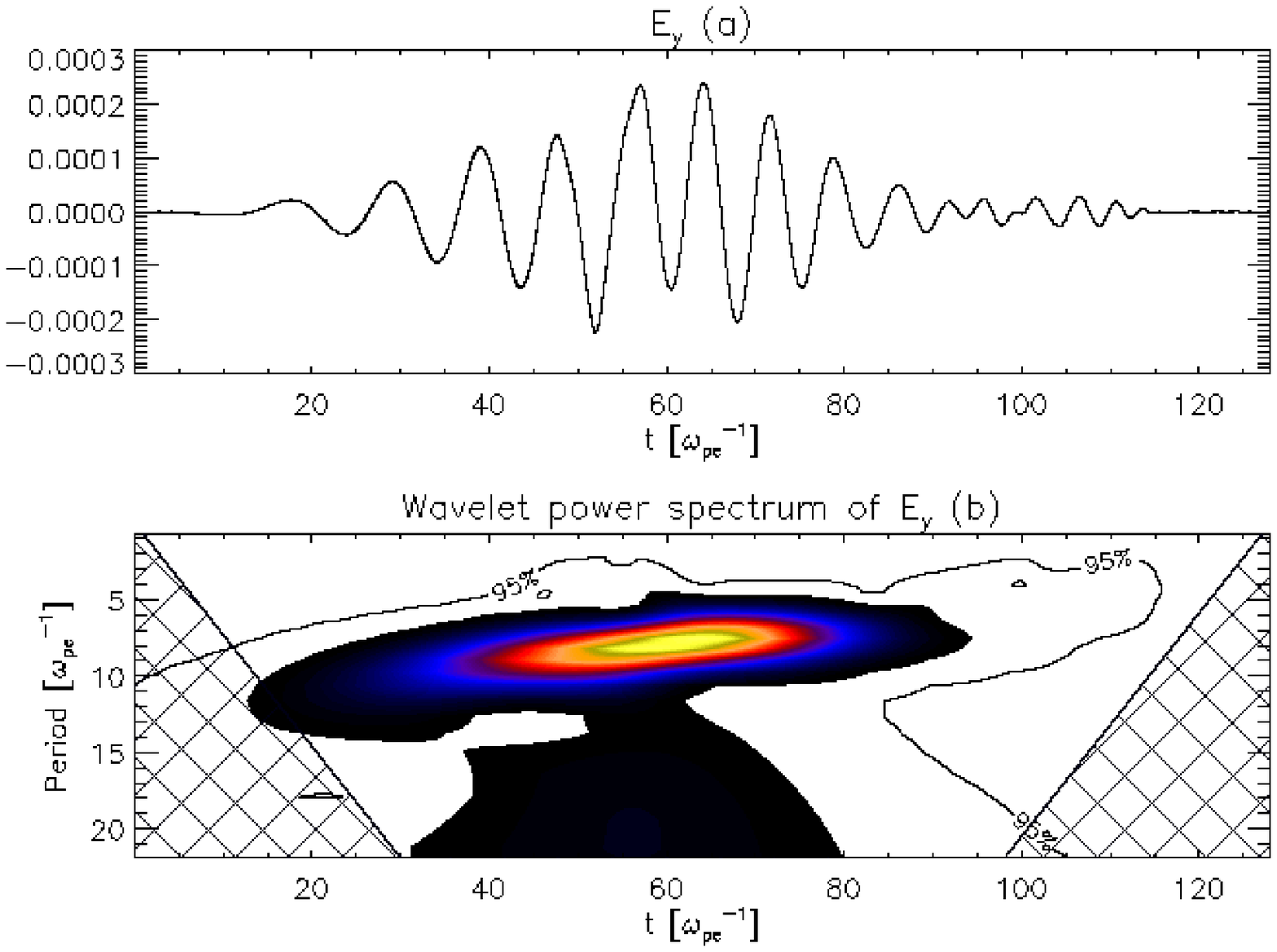}
              }
              \caption{ As in Fig.5 but for 10 times weaker magnetic field such that $\omega_{ce}/\omega_{pe}(x=0)=0.094$.
	      This figure pertains to Section III.e.}
      \end{figure}

\section{Conclusions}

A quote from \cite{melrose99}, p.95, summarises the state of the matters
in theoretical understanding of solar type III radio bursts rather
well: "Our understanding
of plasma emission is in an unsatisfactory state.
It seems that the problems with our understanding of plasma
emission are of an astrophysical nature and will eventually
be solved through new observational data. There are several
different possible mechanisms which can lead to fundamental
plasma emission and it is still not clear which is the
relevant one in practice. Although this leaves the theory
of fundamental plasma emission in a somewhat uncertain
state, the theory for second harmonic is well understood;
there seems to be no reasonable alternative for the coalescence
process $L+L \to T$". Also Introduction section of Ref. \cite{2010JGRA..11501101M} provides a good, 
critical overview of the possible mechanisms which generate
the type III burst EM radiation. These include, (i) the classical
plasma emission mechanism that is based on non-linear wave-wave interaction
between Langmuir, ion-acoustic and EM waves; (ii) a linear mode
conversion, in which almost monochromatic Langmuir z-mode
interacts with the density gradient, partly reflecting and
partly converting into the EM radiation; (iii) the quasimode mechanism in which
forward- and backward- propagating Langmuir waves generate a quasinormal 
electrostatic mode at $2 \omega_{pe}$ which further converts into
EM harmonic radiation and (iv) the antenna radiation model which
involves direct radiation of charged particles that oscillate at $\omega_{pe}$
and drive currents at $2\omega_{pe}$.
In this work we presented the results which show that 
1.5D non-zero pitch angle (non-gyrotropic) electron beam can also
produce escaping EM radiation at $\omega_{pe}$ which seems to
successfully mimic the observed solar type III radio bursts.
A clear distinction needs to be drawn between the best studied type III
burst mechanism, the classical plasma emission \cite{1958SvA.....2..653G},
and the model presented in this work.
In the plasma emission mechanism non-linear wave-wave interaction
between Langmuir, ion-acoustic and EM waves requires
that the beat conditions $\omega_1+\omega_2=\omega$ and
$\vec k_1 +\vec k_2 =\vec k$ to be satisfied.
The {\it emission formula} (i.e.
the three wave interaction probability) for the relevant process 
$L+s \to T$ (coalescence of Langmuir and ion-sound wave to produce transverse
EM wave) includes a cross vector product factor $|\vec k_L \times \vec k_T|^2$
(see e.g. Eqs.(26.24) and (26.25) from Ref.\cite{mmcp2005}.
This implies that, whilst electron beam and Langmuir turbulence
dynamics can be treated in 1D spatial dimensions (and there is large body
of work that deals with the 1D quasilinear theory),
the correct treatment of {\it act of EM emission} needs 2D spatial
dimensions. This is because in 1D case the factor $|\vec k_L \times \vec k_T|^2$
is identically zero, because the angle between wave vectors of
Langmuir {\it wave}, $\vec k_L$, and EM {\it wave}, $\vec k_T$, is zero.
This has to be distinguished from the pitch angle $\theta$ which in
our notation is the angle between the {\it particle} (electron) beam 
injection direction and the background magnetic field.
We remark however that the plasma emission mechanism equations use
"random phase approximation" (see e.g. p.383 from Ref.\cite{mmcp2005}).
This is because the extraneous, quadratic non-linear current (see their Eq.(26.2)) 
depends on the phases of the beating fields and some assumption needs to be made
concerning the phase. {\it A priori}, it is not at all clear that in the case of our
situation, in which non-zero pitch angle (non-gyrotropic) electron beam
is injected, the phases are random. Hence whether 
the factor $|\vec k_L \times \vec k_T|^2$
is applicable in our case. Without further in-depth analysis
it would be safe to conclude that our simulations do not
involve the classical plasma emission processes. 

We have performed-high resolution (sub-Debye length grid size and 10000 particle species per cell), 
1.5D Particle-in-Cell, relativistic, fully electromagnetic simulations 
to model electromagnetic wave emission generation in the context of solar type III radio bursts.
We studied  the generation of EM waves by injecting a super-thermal, hot beam of electrons  into
a plasma thread that contains uniform longitudinal magnetic field and a parabolic density gradient along the magnetic field.
We have considered five cases: 

(i) As an initial equilibrium test, we find that the physical system without electron beam is stable and 
only low amplitude level electromagnetic drift waves (noise) are excited. 

(ii) The beam injection direction is then controlled by
setting either longitudinal or oblique initial electron drift speed/momentum.
i.e. we set different beam pitch angles. 
In the case of zero beam pitch angle, i.e. when $\vec v_b \cdot \vec E_\perp=0$, 
the beam excites only ES standing waves, oscillating at local
plasma frequency. This oscillation occurs strictly in the beam
injection spatial location and only low level electromagnetic drift wave noise is present 
(no regular EM waves are generated by the beam).

(iii) In the case of oblique beam pitch angle, i.e. when $\vec v_b \cdot \vec E_\perp \not =0$ 
again ES waves with similar properties are excited. In this case however
because the beam can interact with the EM waves,
it generates EM waves with the properties commensurate to type III radio bursts. 
In particular, wavelet analysis of transverse electric field component shows that
as the beam moves to the regions of lower density and hence lower plasma frequency,
EM wave frequency drops accordingly. 

(iv)  When we remove the density gradient,  an electron beam with an 
oblique pitch angle still generates the EM radiation, but  now no
frequency decrease is produced. 

(v) In order to prove that the generated, by the
non-zero pitch angle beam, EM emission oscillates at the plasma frequency, we
also consider a case when the magnetic field (and hence 
the cyclotron frequency) is ten times smaller.

Using fully kinetic plasma model our  results also broadly confirm 
(i) the fact that in order to excite escaping EM waves, the electron beam should have non-zero
pitch angle; i.e. there should be a non-zero projection of the 
electron beam injection velocity vector on the transverse
EM electric field vector.
(ii) quasilinear theory predictions, namely quasilinear relaxation time-scale and free streaming assumptions
were corroborated via fully kinetic simulation, in a realistic to the type III burst magnetic field geometry.
(iii) The observational fact that there should be a EM emission
frequency drift in time in the inhomogeneous plasma case has been also confirmed via
production of the simulated (synthetic) dynamical spectrum for the first time.
 
The presented model will be used in the future for the forward modelling of the 
observed type III burst dynamical spectra (2D radio emission intensity plots 
where frequency is on $y$-axis and time on $x$-axis). The main forward modelling goal will be
inferring the electron number density profile along the beam propagation paths.

We would like to close by pointing out some pertinent limitations of the considered model in its direct applicability
to the solar type III radio burst observations. The issues are: 

(i) In the solar type III bursts, electron beams
propagate large distances without being depleted by the generated Langmuir waves due to bump-on-tail 
instability (also called beam-plasma instability). 
In the linear regime, the timescale for the quasilinear relaxation, 
$\tau =n_e/({n_b} \omega_{pe})$ and hence timescale of the beam depletion
is prescribed by the ratio of the electron beam and background plasma densities. 
To be precise, $\tau$ is the inverse of the linear 
bump-on-tail instability growth rate $\gamma = \omega_{pe} (n_b/n_e) (v_b^2/\delta v_b^2)$, 
where $\delta v_b$ is the electron beam velocity thermal spread. 
Thus for $v_b \approx \delta v_b$, $\tau \approx 1/ \gamma$.
Due to the above described computational limitations, at present, it 
was impractical to set $n_b/n_{e}$ to the observed values $10^{-5}-10^{-8}$; 
This may affect the process of beam (re-)generation by the time-of-flight effects; Also, it is known that electron
beam may be stabilised by non-linear effects \cite{ts65}. 
In the non-linear stimulated scattering processes, the wavenumbers are
drawn out of the resonance. This leads to energy transfer rate
between the beam electrons and Langmuir wave at a much slower rate than 
quasilinear relaxation time, that effectively leads to the non-linear 
stabilisation  of the bump-on-tail instability.
For the parameters commensurate to the type III bursts,
the condition for the stabilisation,
$n_b/n_e \ll {v_{th,e}^4}/{v_b^4} \ll (m_e/m_i) (\delta v_b/v_b)$ is met in most cases. Therefore
the non-linear stabilisation is likely to play a major role.
(e.g. Ref.\cite{t95}, pp.184-187 or \cite{ts65}) 

(ii) The spatial scale of the density gradient (decrease of 
$\omega_{pe}$ by a factor of $10^4$ over 65,000 Debye length) 
is not realistic (we have to remember that in PIC simulations one 
uses a smaller than in reality number of "super-particles" and not real 
electrons and protons). This may affect criteria for onset of 
instabilities caused by plasma density inhomogeneities; For example,
when the electron beam moves along the density gradient, Langmuir wave phase velocity
will change while the beam velocity remains constant when there is
no strong relaxation \cite{r70}. If the instability growth rate
is much less than the reciprocal of the time of escape from the resonance, 
the beam stabilises as it no longer loses energy to the wave generation. 
The condition for the stabilisation is $L < (n_e /n_b)(3 v_{th,e}^2/(v_b \omega_{pe}))$, 
where $L$ is the characteristic spatial scale of the plasma density inhomogeneity 
(see e.g. Ref.\cite{kt73}, p. 119). For the broad range of the solar coronal 
conditions as well as for the set of parameters considered in this paper 
(from Figure 4c, thick sold curve, we see that plasma density drops by a 
factor of 2 over a length scale of $L\approx 30 c/\omega_{pe}$, 
whereas $(n_e /n_b)(3 v_{th,e}^2/(v_b \omega_{pe})) = 0.3 c/\omega_{pe}$) the latter inequality is not met. 

(iii) Moreover the observed electron beam pitch angles are also much 
smaller \cite{2009SoPh..259..255R} than considered in the present model. 
Despite these limitations, the present model provides a proof-of-concept 
for the EM emission generation in the context of type III solar radio bursts.

On a positive note, the considered regime may provide an important diagnostic 
to laboratory laser plasma  or thermonuclear fusion studies as in both 
cases non-thermal beams of electrons are frequently present. 

\begin{acknowledgments}
The author would like to thank EPSRC-funded 
Collaborative Computational Plasma Physics  (CCPP) project
lead by Prof. T.D. Arber (Warwick) for providing 
EPOCH Particle-in-Cell code and Dr. K. Bennett (Warwick) for
CCPP related programing support. 
Computational facilities used are that of Astronomy Unit, 
Queen Mary University of London and STFC-funded UKMHD consortium at St Andrews
University. The author is financially supported by HEFCE-funded 
South East Physics Network (SEPNET).
Author would like to thank: Prof. D. Burgess (Queen Mary UL) for useful
discussions and an 
anonymous referee whose comments contributed to an improvement of this paper.
\end{acknowledgments}


\end{document}